\pdfoutput=1

\documentclass[11pt,twoside,a4paper,cmspaper,final,collab]{cms-tdr}
\def\svnVersion{62bf014-D}\def\svnDate{2020/02/29}\def\cmsCernNoTag{CERN-EP-2019-215}\def\cmsCernDate{\today}\def\cmsMessage{"Published in the Journal of High Energy Physics as \href{http://dx.doi.org/10.1007/JHEP04(2020)188}{\doi{10.1007/JHEP04(2020)188}.}"}

\begin{document}\cmsNoteHeader{BPH-16-004}

\hyphenation{had-ron-i-za-tion}
\hyphenation{cal-or-i-me-ter}
\hyphenation{de-vices}

\newcommand{\mup}{\Pgmp}
\newcommand{\mun}{\Pgmm}
\newcommand{\B}{\PB}
\newcommand{\Bs}{\PBzs}
\newcommand{\Bp}{\PBp}
\newcommand{\Bm}{\PBm}
\newcommand{\Bsb}{\PABzs}
\newcommand{\p}{\Pp}
\newcommand{\Dstarp}{\PDstp}
\newcommand{\Dstarm}{\PDstm}
\newcommand{\K}{\PK}
\newcommand{\Kp}{\PKp}
\newcommand{\Km}{\PKm}
\newcommand{\nub}{\Pagngm}
\newcommand{\KS}{\PKzS}
\newcommand{\pip}{\Pgpp}
\newcommand{\pim}{\Pgpm}
\newcommand{\piz}{\Pgpz}
\newcommand{\Dz}{\PDz}
\newcommand{\up}{{\PQu}}
\newcommand{\s}{{\PQs}}
\newcommand{\kstarz}{\PK^{\ast0}}
\newcommand{\ones}{\PgUa}

\newcommand{\BsH}{\ensuremath{{\PB}^0_{\mathrm{sH}}}\xspace}
\newcommand{\BsL}{\ensuremath{{\PB}^0_{\mathrm{sL}}}\xspace}
\newcommand{\BsLH}{\ensuremath{{\PB}^0_{\mathrm{sL(H)}}}\xspace}
\newcommand{\fsfu}{\ensuremath{f_{\PQs}/f_{\PQu}}\xspace}
\newcommand{\fsfd}{\ensuremath{f_{\PQs}/f_{\PQd}}\xspace}
\newcommand{\fufs}{\ensuremath{f_{\PQu}/f_{\PQs}}\xspace}
\newcommand{\fdfu}{\ensuremath{f_{\PQd}/f_{\PQu}}\xspace}
\newcommand{\tmm}{ \ensuremath{\tau_{\mup\mun}}\xspace }
\newcommand{\adgmm}{\ensuremath{{\mathcal{A}}_{\Delta\Gamma}^{\mup\mun}}\kern-0.2em}
\newcommand{\mll}{\ensuremath{m_{\mup\mun}}\xspace}
\newcommand{\etaf}{\ensuremath{\eta^{\text{f}}_{\mu}}\xspace}
\newcommand{\dca}{\ensuremath{d_{\text{ca}}}\xspace}
\newcommand{\fls}{\ensuremath{\ell_{\mathrm{3D}}/\sigma(\ell_{\mathrm{3D}})}\xspace}
\newcommand{\flsxy}{\ensuremath{\ell_{xy}/\sigma(\ell_{xy})}\xspace}
\newcommand{\pvip}{\ensuremath{\delta_{\mathrm{3D}}}\xspace}
\newcommand{\pvips}{\ensuremath{\delta_{\mathrm{3D}}/\sigma(\delta_{\mathrm{3D}})}\xspace}
\newcommand{\fl}{\ensuremath{\ell_{\mathrm{3D}}}\xspace}

\newcommand{\fle}{\ensuremath{\sigma(\ell_{\mathrm{3D}})}\xspace}
\newcommand{\ptb}{\ensuremath{{\pt}_{\B}}\xspace}
\newcommand{\ptmu}{\ensuremath{{\pt}_{\mu}}\xspace}
\newcommand{\ptmm}{\ensuremath{{\pt}_{\mup\mun}}\xspace}
\newcommand{\closetrk}{\ensuremath{N_{\text{trk}}^{\text{close}}}\xspace}
\newcommand{\docatrk}{\ensuremath{d^{\text{0}}_{\text{ca}}}\xspace}
\newcommand{\maxdoca}{\ensuremath{d_{\text{ca}}^{\text{max}}}\xspace}
\newcommand{\mkk}{\ensuremath{m_{\Kp\Km}}\xspace}
\newcommand{\pa}{\ensuremath{\alpha_{\mathrm{3D}}}\xspace}
\newcommand{\iso}{\ensuremath{I}\xspace}
\newcommand{\isomuone}{\ensuremath{I_{\mu_1}}\xspace}
\newcommand{\isomutwo}{\ensuremath{I_{\mu_2}}\xspace}
\newcommand{\st}{\ensuremath{\sigma_t}\xspace}
\newcommand{\ps}{\unit{ps}}
\newcommand{\invps}{\mbox{\ensuremath{\,\text{ps}^{-1}}}\xspace}
\newcommand{\cbf}{\ensuremath{\mathcal{B}}\xspace}
\newcommand{\calc}{\ensuremath{\mathcal{C}}\xspace}
\newcommand{\calP}{\ensuremath{\mathcal{P}}\xspace}
\newcommand{\chidof}{\ensuremath{\chi^2/\text{dof}}\xspace}
\newcommand{\bmm}{\ensuremath{\PB\to\MM}\xspace}
\newcommand{\bdmm}{\ensuremath{\PBz\to\MM}\xspace}
\newcommand{\bsmm}{\ensuremath{\PBzs\to\MM}\xspace}
\newcommand{\bsdmm}{\ensuremath{{\PB}^0_{{\PQs,\PQd}}\to\MM}\xspace}
\newcommand{\bsdll}{\ensuremath{{\PB}^0_{{\PQs,\PQd}}\to\ell^+\ell^-}\xspace}
\newcommand{\bqmm}{\ensuremath{{\PB}^0_{{\PQq}}\to\MM}\xspace}
\newcommand{\jpsi}{\ensuremath{\PJGy}\xspace}
\newcommand{\bupsikp}{\ensuremath{\PBp\to\jpsi\PKp}\xspace}
\newcommand{\bupsipip}{\ensuremath{\PBp\to\jpsi\PGpp}\xspace}
\newcommand{\bdpsikstarz}{\ensuremath{\PBz\to\jpsi\PKst^0}\xspace}
\newcommand{\bcpsimunu}{\ensuremath{\PBpc\to\jpsi\PGmp\PGn}\xspace}
\newcommand{\bzspsiphi}{\ensuremath{\PBzs\to\jpsi\PGf}\xspace}
\newcommand{\BsBsb}{\ensuremath{\PBzs\text{--}\PABzs}\xspace}
\newcommand{\h}{{\HepParticle{h}{}{}}\xspace}
\newcommand\hprime{\ensuremath{\h^{(\raisebox{-1ex}{'})}}\xspace}
\newcommand{\bhm}{\ensuremath{\PB\to\h\mu\nu}\xspace}
\newcommand{\bhmm}{\ensuremath{\PB\to\h\mu\mu}\xspace}
\newcommand{\bhh}{\ensuremath{\PB\to\h\hprime}\xspace}
\newcommand{\ys}{\ensuremath{y_{\mathrm{s}}}\xspace}
\newcommand{\Gs}{\ensuremath{\Gamma_{\mathrm{s}}}\xspace}

\newcommand{\StrutTiny}{\rule[-2.mm]{0pt}{6mm}}
\definecolor{dgreen}{rgb}{0.,0.5,0.}
\newlength\cmsTabSkip\setlength{\cmsTabSkip}{1ex}
\providecommand{\NA}{\ensuremath{\text{---}}\xspace}

\newcommand*\xbar[1]{%
  \hbox{%
    \vbox{%
      \hrule height 0.4pt%
      \kern0.5ex
      \hbox{%
        \kern-0.em
        \ensuremath{#1}%
        \kern-0.em
      }%
    }%
    \raisebox{9.0pt}{\scalebox{.4}{\text{\kern-2.3em(}}} %
    \raisebox{9.0pt}{\scalebox{.4}{\text{\kern1.3em)}}} %
  }%
}

\newcommand*\xbarr[1]{%
   \hbox{%
     \vbox{%
       \hrule height 0.4pt%
       \kern0.5ex
       \hbox{%
         \kern0.0em
         \ensuremath{#1}%
         \kern0.0em
       }%
     }%
     \raisebox{9.0pt}{\scalebox{.4}{\text{\kern-1.9em(}}} %
     \raisebox{9.0pt}{\scalebox{.4}{\text{\kern0.9em)}}} %
   }%
}

\newcommand\vdef[1]{\expandafter\def\csname #1\endcsname}
\newcommand\vuse[1]{\csname #1\endcsname}
\newcommand\vu[1]{\csname #1\endcsname}

\newcommand{\resObsBFBsmm}{\ensuremath{[2.9\pm{0.7}\,(\text{exp})\pm{0.2}\,(\text{frag}) ] \times10^{-9}}}
\newcommand{\resObsBFBsmmLong}{\ensuremath{[2.9\pm{0.6}\stat \pm{0.3}\syst\pm{0.2}\,(\text{frag}) ] \times10^{-9}}}
\newcommand{\resObsBFBdmm}{\ensuremath{(0.8\,^{+1.4}_{-1.3})\times10^{-10}}}
\newcommand{\resObsTauBsmm}{\ensuremath{1.70\,^{+0.61}_{-0.44}}}
\newcommand{\resObsTauBsmmLong}{\ensuremath{[1.70\,^{+0.60}_{-0.43}\stat\pm{0.09}\syst]}}
\newcommand{\sigObsBFBsmm}{\ensuremath{5.6}}
\newcommand{\sigExpBFBsmm}{\ensuremath{6.5}}
\newcommand{\sigObsBFBdmm}{\ensuremath{0.6}}
\newcommand{\sigExpBFBdmm}{\ensuremath{0.8}}
\newcommand{\ulaBFBdmm}{\ensuremath{3.6\times10^{-10}}}
\newcommand{\ulbBFBdmm}{\ensuremath{3.1\times10^{-10}}}
\newcommand{\ulacl}{\ensuremath{95}}
\newcommand{\ulbcl}{\ensuremath{90}}
\newcommand{\ulaExpBdmm}{\ensuremath{3.0\times10^{-10}}}
\newcommand{\ulbExpBdmm}{\ensuremath{2.4\times10^{-10}}}
\newcommand{\resObsTauSplot}{\ensuremath{1.55\ ^{+0.52}_{-0.33}}}
\newcommand{\resObsBFBsmmRunA}{\ensuremath{(2.3\ ^{+1.0}_{-0.8})\times10^{-9}}}
\newcommand{\sigObsBFBsmmRunA }{\ensuremath{3.3}}
\newcommand{\sigExpBFBsmmRunA}{\ensuremath{4.5}}

\vdef{cat_2011s01_0_0_N_Bs} {\ensuremath{3.6\ ^{+0.9}_{-0.8} } }
\vdef{cat_2011s01_0_0_N_Bd} {\ensuremath{0.4\ ^{+0.7}_{-0.6} } }
\vdef{cat_2011s01_0_0_N_comb} {\ensuremath{2.3\pm1.0 } }
\vdef{cat_2011s01_0_0_N_Bu} {\ensuremath{750\pm30 } }
\vdef{cat_2011s01_0_0_Eff_Bs_Bu} {\ensuremath{3.9\pm0.5 } }
\vdef{cat_2011s01_1_0_N_Bs} {\ensuremath{2.0\ ^{+0.5}_{-0.4} } }
\vdef{cat_2011s01_1_0_N_Bd} {\ensuremath{0.2\ ^{+0.4}_{-0.3} } }
\vdef{cat_2011s01_1_0_N_comb} {\ensuremath{0.7\pm0.5 } }
\vdef{cat_2011s01_1_0_N_Bu} {\ensuremath{220\pm10 } }
\vdef{cat_2011s01_1_0_Eff_Bs_Bu} {\ensuremath{7.5\pm0.8 } }
\vdef{cat_2012s01_0_0_N_Bs} {\ensuremath{3.7\ ^{+0.9}_{-0.8} } }
\vdef{cat_2012s01_0_0_N_Bd} {\ensuremath{0.4\ ^{+0.6}_{-0.6} } }
\vdef{cat_2012s01_0_0_N_comb} {\ensuremath{29.9\pm2.9 } }
\vdef{cat_2012s01_0_0_N_Bu} {\ensuremath{790\pm30 } }
\vdef{cat_2012s01_0_0_Eff_Bs_Bu} {\ensuremath{3.8\pm0.5 } }
\vdef{cat_2012s01_0_1_N_Bs} {\ensuremath{9.3\ ^{+2.3}_{-2.1} } }
\vdef{cat_2012s01_0_1_N_Bd} {\ensuremath{1.0\ ^{+1.7}_{-1.6} } }
\vdef{cat_2012s01_0_1_N_comb} {\ensuremath{7.6\pm1.8 } }
\vdef{cat_2012s01_0_1_N_Bu} {\ensuremath{2360\pm100 } }
\vdef{cat_2012s01_0_1_Eff_Bs_Bu} {\ensuremath{3.2\pm0.4 } }
\vdef{cat_2012s01_1_0_N_Bs} {\ensuremath{1.7\ ^{+0.4}_{-0.4} } }
\vdef{cat_2012s01_1_0_N_Bd} {\ensuremath{0.2\ ^{+0.3}_{-0.3} } }
\vdef{cat_2012s01_1_0_N_comb} {\ensuremath{29.9\pm2.9 } }
\vdef{cat_2012s01_1_0_N_Bu} {\ensuremath{190\pm10 } }
\vdef{cat_2012s01_1_0_Eff_Bs_Bu} {\ensuremath{7.3\pm1.0 } }
\vdef{cat_2012s01_1_1_N_Bs} {\ensuremath{4.7\ ^{+1.2}_{-1.1} } }
\vdef{cat_2012s01_1_1_N_Bd} {\ensuremath{0.5\ ^{+0.9}_{-0.8} } }
\vdef{cat_2012s01_1_1_N_comb} {\ensuremath{8.3\pm1.7 } }
\vdef{cat_2012s01_1_1_N_Bu} {\ensuremath{660\pm30 } }
\vdef{cat_2012s01_1_1_Eff_Bs_Bu} {\ensuremath{5.9\pm0.8 } }
\vdef{cat_2016BFs01_0_0_N_Bs} {\ensuremath{2.2\ ^{+0.5}_{-0.5} } }
\vdef{cat_2016BFs01_0_0_N_Bd} {\ensuremath{0.2\ ^{+0.4}_{-0.4} } }
\vdef{cat_2016BFs01_0_0_N_comb} {\ensuremath{10.3\pm1.7 } }
\vdef{cat_2016BFs01_0_0_N_Bu} {\ensuremath{580\pm20 } }
\vdef{cat_2016BFs01_0_0_Eff_Bs_Bu} {\ensuremath{3.1\pm0.4 } }
\vdef{cat_2016BFs01_0_1_N_Bs} {\ensuremath{4.0\ ^{+1.0}_{-0.9} } }
\vdef{cat_2016BFs01_0_1_N_Bd} {\ensuremath{0.4\ ^{+0.8}_{-0.7} } }
\vdef{cat_2016BFs01_0_1_N_comb} {\ensuremath{3.4\pm1.2 } }
\vdef{cat_2016BFs01_0_1_N_Bu} {\ensuremath{1290\pm60 } }
\vdef{cat_2016BFs01_0_1_Eff_Bs_Bu} {\ensuremath{2.5\pm0.3 } }
\vdef{cat_2016BFs01_1_0_N_Bs} {\ensuremath{3.7\ ^{+0.9}_{-0.8} } }
\vdef{cat_2016BFs01_1_0_N_Bd} {\ensuremath{0.4\ ^{+0.7}_{-0.7} } }
\vdef{cat_2016BFs01_1_0_N_comb} {\ensuremath{43.5\pm3.5 } }
\vdef{cat_2016BFs01_1_0_N_Bu} {\ensuremath{780\pm30 } }
\vdef{cat_2016BFs01_1_0_Eff_Bs_Bu} {\ensuremath{3.9\pm0.5 } }
\vdef{cat_2016BFs01_1_1_N_Bs} {\ensuremath{8.1\ ^{+2.0}_{-1.8} } }
\vdef{cat_2016BFs01_1_1_N_Bd} {\ensuremath{0.8\ ^{+1.5}_{-1.4} } }
\vdef{cat_2016BFs01_1_1_N_comb} {\ensuremath{15.9\pm2.4 } }
\vdef{cat_2016BFs01_1_1_N_Bu} {\ensuremath{1920\pm80 } }
\vdef{cat_2016BFs01_1_1_Eff_Bs_Bu} {\ensuremath{3.4\pm0.4 } }
\vdef{cat_2016GHs01_0_0_N_Bs} {\ensuremath{4.1\ ^{+1.0}_{-0.9} } }
\vdef{cat_2016GHs01_0_0_N_Bd} {\ensuremath{0.4\ ^{+0.8}_{-0.7} } }
\vdef{cat_2016GHs01_0_0_N_comb} {\ensuremath{34.4\pm3.2 } }
\vdef{cat_2016GHs01_0_0_N_Bu} {\ensuremath{1020\pm40 } }
\vdef{cat_2016GHs01_0_0_Eff_Bs_Bu} {\ensuremath{3.3\pm0.4 } }
\vdef{cat_2016GHs01_0_1_N_Bs} {\ensuremath{3.6\ ^{+0.9}_{-0.8} } }
\vdef{cat_2016GHs01_0_1_N_Bd} {\ensuremath{0.4\ ^{+0.7}_{-0.6} } }
\vdef{cat_2016GHs01_0_1_N_comb} {\ensuremath{2.2\pm1.0 } }
\vdef{cat_2016GHs01_0_1_N_Bu} {\ensuremath{1320\pm50 } }
\vdef{cat_2016GHs01_0_1_Eff_Bs_Bu} {\ensuremath{2.2\pm0.2 } }
\vdef{cat_2016GHs01_1_0_N_Bs} {\ensuremath{6.1\ ^{+1.5}_{-1.4} } }
\vdef{cat_2016GHs01_1_0_N_Bd} {\ensuremath{0.6\ ^{+1.1}_{-1.0} } }
\vdef{cat_2016GHs01_1_0_N_comb} {\ensuremath{33.4\pm3.1 } }
\vdef{cat_2016GHs01_1_0_N_Bu} {\ensuremath{1260\pm50 } }
\vdef{cat_2016GHs01_1_0_Eff_Bs_Bu} {\ensuremath{3.9\pm0.4 } }
\vdef{cat_2016GHs01_1_1_N_Bs} {\ensuremath{3.9\ ^{+1.0}_{-0.9} } }
\vdef{cat_2016GHs01_1_1_N_Bd} {\ensuremath{0.4\ ^{+0.8}_{-0.7} } }
\vdef{cat_2016GHs01_1_1_N_comb} {\ensuremath{4.0\pm1.3 } }
\vdef{cat_2016GHs01_1_1_N_Bu} {\ensuremath{1180\pm50 } }
\vdef{cat_2016GHs01_1_1_Eff_Bs_Bu} {\ensuremath{2.7\pm0.3 } }
\vdef{cat_2011s01_0_0_ptBs} {\ensuremath{16.4 } }
\vdef{cat_2011s01_1_0_ptBs} {\ensuremath{14.9 } }
\vdef{cat_2012s01_0_0_ptBs} {\ensuremath{16.1 } }
\vdef{cat_2012s01_0_1_ptBs} {\ensuremath{17.3 } }
\vdef{cat_2012s01_1_0_ptBs} {\ensuremath{14.3 } }
\vdef{cat_2012s01_1_1_ptBs} {\ensuremath{15.5 } }
\vdef{cat_2016BFs01_0_0_ptBs} {\ensuremath{17.5 } }
\vdef{cat_2016BFs01_0_1_ptBs} {\ensuremath{19.3 } }
\vdef{cat_2016BFs01_1_0_ptBs} {\ensuremath{15.8 } }
\vdef{cat_2016BFs01_1_1_ptBs} {\ensuremath{17.5 } }
\vdef{cat_2016GHs01_0_0_ptBs} {\ensuremath{17.2 } }
\vdef{cat_2016GHs01_0_1_ptBs} {\ensuremath{20.8 } }
\vdef{cat_2016GHs01_1_0_ptBs} {\ensuremath{16.2 } }
\vdef{cat_2016GHs01_1_1_ptBs} {\ensuremath{19.5 } }
\vdef{sum_N_Bs} {\ensuremath{61\ ^{+15}_{-13}}}
\vdef{sum_N_Bu_text} {\ensuremath{(1.43\pm0.06)\times10^6}}

\cmsNoteHeader{BPH-16-004}

\title{Measurement of properties of \bsmm decays and search for \bdmm with the CMS experiment}

\date{\today}

\abstract{Results are reported for the \bsmm\ branching fraction and effective lifetime and from a search
  for the decay \bdmm.  The analysis uses a data sample of
  proton-proton collisions accumulated by the CMS experiment in 2011,
  2012, and 2016, with center-of-mass energies (integrated
  luminosities) of 7\TeV\ (5\fbinv), 8\TeV\ (20\fbinv), and 13\TeV\
  (36\fbinv).  The branching fractions are determined by measuring
  event yields relative to \bupsikp\ decays (with
  $\jpsi\to\mup\mun$), which results in the reduction of many of
  the systematic uncertainties. The decay \bsmm\ is observed with
  a significance of 5.6~standard deviations. The
  branching fraction is measured to be $\mathcal{B}(\bsmm) =
  [2.9\ \pm0.7(\text{exp})\pm0.2(\text{frag})]\times 10^{-9}$,
  where the first uncertainty combines the experimental
  statistical and systematic contributions, and the
  second is due to the uncertainty in the ratio
  of the \PBzs\ and the \PBp\ fragmentation functions.  No
  significant excess is observed for the decay \bdmm, and an upper
  limit of $\cbf(\bdmm) < 3.6\times10^{-10}$ is obtained at 95\% confidence
  level. The \bsmm\
  effective lifetime is measured to be $\tmm = 1.70\,^{+0.61}_{-0.44}\ps$.
  These results are consistent with standard model predictions.  }

\hypersetup{%
pdfauthor={CMS Collaboration},%
pdftitle={Measurement of properties of Bs0 to mu+mu- decays and search for
B0 to mu+mu- with the CMS experiment},%
pdfsubject={CMS},%
pdfkeywords={CMS, B physics}}

\maketitle

\section{Introduction}
\label{s:intro}
Leptonic \PB meson decays offer excellent opportunities to perform
precision tests of the standard model (SM) of particle physics because
of minimal hadronic uncertainties in the theoretical
predictions~\cite{Bobeth:2013uxa,Bobeth:2013tba,Hermann:2013kca,Beneke:2017vpq,Beneke:2019slt}.
In the SM, the decays \bsmm\ and \bdmm proceed only via loop diagrams and are
also helicity suppressed, leading to very small expected decay
time-integrated branching fractions,
$(3.66\pm0.14)\times 10^{-9}$ and $(1.03\pm0.05)\times 10^{-10}$,
respectively.
Theoretical uncertainties in the calculation of these branching
fractions have been reduced in recent years as a result of progress in
lattice quantum chromodynamics (QCD)~\cite{Aoki:2019cca,
  Bazavov:2017lyh, Bussone:2016iua, Dowdall:2013tga, Hughes:2017spc},
in the calculation of electroweak effects
at next-to-leading order~\cite{Bobeth:2013tba}, and in the calculation of
QCD effects at next-to-next-to-leading order~\cite{Hermann:2013kca}.
Enhanced electromagnetic contributions from virtual photon exchange
have been shown~\cite{Beneke:2017vpq, Beneke:2019slt} to produce larger corrections
than previously assumed in the theoretical uncertainties.
The \bsmm\ branching fraction has been measured  in proton-proton ($\Pp\Pp$)
collisions by the CMS, LHCb, and ATLAS
Collaborations~\cite{Chatrchyan:2013bka,CMS:2014xfa,Aaij:2017vad,Aaboud:2018mst}.
For the decay \bdmm, evidence at the three standard deviation level has
been obtained by the CMS and LHCb Collaborations in a combined
analysis~\cite{CMS:2014xfa} of $\sqrt{s} = 7$ and 8\TeV data. However, this has not been confirmed by
LHCb after incorporating 13\TeV data~\cite{Aaij:2017vad}, nor by ATLAS~\cite{Aaboud:2018mst}.

The heavy (\BsH) and light (\BsL) mass eigenstates are linear
combinations of the flavor eigenstates,
$|\BsLH\rangle = p|\Bs\rangle \pm q|\Bsb\rangle$, with the
normalization condition $\abs{p}^2
+ \abs{q}^2 = 1$. The decay amplitudes
$\xbar{A}_{\kern-0.4em\mup\mun} = A(\xbarr{\PB}_{\kern-0.4em\mathrm{s}}{}^{\kern-0.2em0}\to\mup\mun)$,
together with $p$ and $q$, are used to
define~\cite{Amhis:2016xyh} the parameters $\lambda \equiv
(q/p)(\overline{A}_{\mup\mun}/A_{\mup\mun})$ and $\adgmm \equiv
-2\Re(\lambda)/(1+\abs{\lambda}^2)$.
The SM predicts $\adgmm = +1$, \ie, that only the heavy state, with measured lifetime
$\tau_{\BsH} = 1.615\pm0.009\ps$~\cite{pdg2018}, contributes
to the \bsmm\ decay. The \bsmm\ effective lifetime is defined by
\begin{linenomath}
  \begin{equation}
    \tmm \equiv \frac
         {\int_{0}^{\infty} t\,[\Gamma(\Bs(t) \to\mup\mun)+\Gamma(\Bsb(t) \to\mup\mun)]\,\rd{}t}
         {\int_{0}^{\infty} [\Gamma(\Bs(t) \to\mup\mun)+\Gamma(\Bsb(t) \to\mup\mun)]\,\rd{}t},
  \end{equation}
\end{linenomath}
where $t$ is the proper decay time of the \Bs\ meson~\cite{DeBruyn:2012wk}. The
effective lifetime is related to the \Bs\ mean lifetime through~\cite{DeBruyn:2012wj}
\begin{linenomath}
  \begin{equation}
    \tmm = \frac{\tau_{\Bs}}{1-\ys^2}\left(\frac{1+2\adgmm\, \ys +  \ys^2}{1+\adgmm\, \ys}\right),
  \end{equation}
\end{linenomath}
where the parameter $\ys\equiv
\tau_{\Bs}\Delta\Gs/2 = 0.066\pm0.004$ is defined using the
measured \Bs\ mean lifetime $\tau_{\Bs} = 1.509\pm0.004\ps$ and
\BsBsb\ decay width difference
$\Delta \Gs \equiv {\Gs}_{\text{L}} - {\Gs}_{\text{H}} = 0.088\pm 0.006 \invps$~\cite{pdg2018}.
A first measurement of the effective lifetime,
$\tmm =2.04\pm0.44\pm0.05\ps$, which is consistent with the SM expectation,
has been presented by the LHCb Collaboration~\cite{Aaij:2017vad}.

In this paper, we report updated results for the \bsmm\ and
\bdmm\ branching fractions, as well as a measurement of the
\bsmm\ effective lifetime. The data
were collected in  $\Pp\Pp$ collisions at the CERN LHC,
and correspond to integrated luminosities of 5 and 20\fbinv
recorded in 2011 and 2012 at $\sqrt{s} = 7$ and 8\TeV,
respectively, during Run~1 of the LHC, and 36\fbinv recorded in 2016 at
$\sqrt{s}=13\TeV$, during Run~2. Depending on the context, the symbol \PB is used to denote
the \Bz, \Bs, and \Bp\ mesons and/or the \PGLb baryon, and charge conjugation
is implied throughout, except as noted.
The present branching fraction measurements supersede the
previous CMS results~\cite{Chatrchyan:2013bka}, which used
Run~1 data only. The main differences with respect to the previous
branching fraction results include the greater statistical precision of the larger data sample,
an improved muon identification algorithm, based on a newly developed boosted decision tree (BDT), and
better constraints against background contamination in the search for
\bdmm.
In addition to the BDT used for muon identification, the analysis employs a second BDT in
the candidate selection. For clarity we will refer to them as the muon BDT and the analysis BDT.
The Run~1 data are reanalyzed using the new muon
identification algorithm (with its improved BDT), but the candidate
selection incorporates the same analysis BDT as used in the original Run~1
analysis, as described in Ref.~\cite{Chatrchyan:2013bka}.
The binning of the analysis BDT discriminator distribution, used for the final
result extraction, was modified, based on the best expected
performance. For the Run~2 data, a new analysis BDT was developed.

The signal sample consists of \PB candidates constructed from two
oppositely charged muons, which are constrained to originate from a
common origin and have an invariant mass in the range $4.8 < \mll <
6.0\GeV$. Within the signal sample, a signal region defined by
$5.20 < \mll <5.45 \GeV$ is analyzed only after all analysis
procedures have been finalized.
The background is estimated from mass sidebands in data and from Monte
Carlo (MC) simulation for specific background sources from \PB
decays. The main background categories are (1) combinatorial
background with two genuine muons from semileptonic decays of separate
\PB hadrons (\eg,
$\Bz\to\Dstarm\mup\nu$), (2) rare \PB decays with two muons (\eg,
from \bhmm\ where $\h\in\{\PGp, \PK, \Pp\}$), and (3) rare \PB decays
with one hadron (\eg,
from \bhm) or two hadrons (\eg, from \bhh)
misidentified as muons.  The combinatorial background affects
both \bsmm\ and \bdmm\ and is the limiting factor for the measurement
of the former. The search for the decay \bdmm, with its smaller expected
branching fraction and an expected signal-to-background ratio
significantly below one, is additionally affected by rare \PB decays, since background
from hadronic $\B$ decays produces a dimuon invariant mass distribution that peaks underneath the \bdmm\ signal.
The background from rare \PB decays has only a minor impact on
the \bsmm\ results.

Because the mass resolution of the CMS detector has a strong
dependence on the pseudorapidity $\eta$ of the muons, the analysis
sensitivity benefits from a division of the data sets into channels
based on the pseudorapidity $\etaf$ of the most forward muon of the
\PB candidate, where $\abs{\etaf}
= \max(\abs{\eta_{\mup}}, \abs{\eta_{\mun}})$. A central and a forward
channel are defined for all running periods, with different boundaries for
Run~1 and Run~2 because of changing trigger requirements.

A normalization sample based on \bupsikp\ decays (with
$\jpsi\to\mup\mun$) is used in the measurement of the branching fractions. In
addition, a control sample based on \bzspsiphi\ decays (with
$\jpsi\to\mup\mun$ and $\phi\to\Kp\Km$) is used to study differences between \Bp\ and \Bs\
characteristics (fragmentation, isolation, selection efficiency, etc.)
in data and to compare with MC simulation. These samples are
reconstructed by adding one or two charged tracks with a kaon mass
hypothesis to two oppositely charged muons, requiring the dimuon pair
to be consistent with \jpsi\ meson decay.

The \bsmm\ branching fraction is determined using
\begin{linenomath}
  \begin{equation}
    \cbf(\bsmm)
    =  \frac{N_\mathrm{S}}{N_{\text{obs}}^{\Bp}} \,
    \frac{f_{\up}}{f_{\s}} \,
    \frac{\varepsilon_{\text{tot}}^{\Bp}}{\varepsilon_{\text{tot}}} \,
    \cbf(\Bp\to\jpsi\Kp)\, \cbf(\jpsi\to\mup\mun)\label{eq:schema},
  \end{equation}
\end{linenomath}
where $N_\mathrm{S}$
($N_{\text{obs}}^{\Bp}$) is the number of reconstructed
\bsmm\ (\bupsikp) decays, $\varepsilon_\text{tot}$
($\varepsilon_{\text{tot}}^{\Bp}$) is the total signal (\Bp)
efficiency, $\cbf(\Bp\to\jpsi\Kp) = (1.01\pm0.03)\times10^{-3}$ and
$\cbf(\jpsi\to\mup\mun) = (5.96\pm0.03)\times10^{-2}$~\cite{pdg2018},
and $\fufs$ is the ratio of the \Bp\ and \Bs\ fragmentation functions.
The value $\fsfu=0.252\pm0.012\,\text{(exp)}\pm0.015\,\text{(CMS)}$, a
combination~\cite{pdg2018} with input from measurements by the
LHCb~\cite{Aaij:2013qqa} and ATLAS Collaborations~\cite{Aad:2015cda},
is used.  Beyond the experimental uncertainty from
Ref.~\cite{pdg2018}, we assign an additional uncertainty
(labeled CMS) by adding in quadrature uncertainties evaluated from the
consideration of two other issues.  First, we derive an uncertainty of
0.008 from the difference between the value of \fsfu\ in
Ref.~\cite{pdg2018}, obtained at $\sqrt{s}=7\TeV$, and that in
Ref.~\cite{Aaij:2019pqz}, obtained at $\sqrt{s}=13\TeV$.
Second, using the parametrization of the
transverse momentum (\pt) dependence in Ref.~\cite{Aaij:2019pqz}, we
determine a difference of 0.013 between the \fsfu\ values at the
average \pt\ of Ref.~\cite{Aaij:2019pqz} and the average \pt\ of
the \bsmm\ candidates in this analysis (see below,
Table~\ref{t:bfyields}). This also covers the substantially smaller
\pt\ dependence of Ref.~\cite{Aaij:2019eej}.
An analogous equation to Eq.~(\ref{eq:schema}) is used to determine
the \bdmm\ branching fraction, where we assume $\fdfu = 1$ for the
ratio of the \Bz\ to the \Bp\ fragmentation functions~\cite{pdg2018}.

The measurement of $\cbf(\bsmm)$ and
$\cbf(\bdmm)$ is performed with an extended unbinned maximum likelihood (UML)
fit, with probability density functions (PDFs) obtained from simulated
event samples and data sidebands. For the determination of the
\bsmm\ effective lifetime $\tmm$, two independent procedures are
used. The first is based on a two-dimensional (2D) UML fit to the
invariant mass and proper decay time distributions of
\bsmm\ candidates. The second is based on a one-dimensional (1D) binned maximum
likelihood (ML) fit to the background-subtracted proper decay time
distribution obtained with the \textit{sPlot}~\cite{Pivk:2004ty}  method.

The presence of multiple $\Pp\Pp$ interactions in an event is referred to as
pileup, whose rate is dependent on the instantaneous luminosity.  The
average number of reconstructed $\Pp\Pp$ interaction vertices is 8, 15, and
18 for the data collected in 2011, 2012, and 2016, respectively.

\section{Event simulation}
\label{s:mc}
Simulated event samples, produced with MC programs, are used to
optimize the analysis selection requirements and to determine
efficiencies for the signal, normalization, and control
samples. The background shapes of the dimuon invariant
mass distribution for rare \PB decays, where one or two charged
hadrons are misidentified as muons, are also obtained from simulated event
samples. The simulated background decay modes include \bhm, \bhmm, and \bhh. For the decay $\PGLb\to \p \mun \nub$, the
model of Ref.~\cite{Khodjamirian:2011jp}, based on QCD light-cone sum
rules, is used. In addition, the decay \bcpsimunu\ was studied, but is
not required for an adequate background description after the full
selection has been applied and therefore is not considered in the
final analysis.

The simulated event samples are generated with
\PYTHIA\ 6.426~\cite{Sjostrand:2006za} for the Run~1 analysis and
\PYTHIA\ 8.212~\cite{Sjostrand:2014zea} for the Run~2 analysis. In
both cases, signal and background events are selected from generic
$2\to2$ QCD processes to provide a complete mixture of gluon fusion,
gluon splitting, and flavor excitation production.  The analysis
efficiency varies for these production mechanisms. For instance, in
gluon splitting the two \PQb quarks can have a small phase space
separation such that a \bsmm\ decay is less isolated than in gluon fusion,
for which the two \PQb quarks are, to first order, back-to-back in the
transverse plane. The mixture in the MC simulation is compared to the
mixture in data using the normalization and control samples, and a
corresponding systematic uncertainty is assigned. The changing pileup
conditions are reflected in the event simulation.

The decay of unstable particles is described using the \EVTGEN~\cite{Lange:2001uf}
program and final-state photon radiation using the
\PHOTOS~\cite{Golonka:2005pn,Davidson:2010ew} program.  The detector
response is simulated with \GEANTfour~\cite{Agostinelli:2002hh}.

\section{The CMS detector}
\label{s:cms}
The CMS experiment is based on a general purpose detector designed and built to
study physics at the TeV scale.
A detailed description of the CMS detector, together with a definition
of the coordinate system used and the relevant kinematic variables,
can be found in Ref.~\cite{Chatrchyan:2008zzk}.  For this analysis,
the main subdetectors used are the silicon tracker,
composed of pixel and microstrip detectors within a 3.8\unit{T} axial magnetic
field, and the muon detector, described below. These detectors are divided into a barrel and two
endcap sections.

The silicon tracker detects charged particles within $\abs{\eta} <
2.5$. The pixel detector is composed of three layers in the barrel
region and two disks located on each side in the forward regions.  In
total, the pixel detector contains about 66 million
$100\mum\times150\mum$ pixels.  Further from the interaction region is
a microstrip detector, composed of ten barrel layers, and three inner
and nine outer disks on either end of the detector, with strip pitches
between 80 and 180$\mum$.  In total, the microstrip detector contains
around 10 million strips and, together with the pixel detector,
yields an impact parameter resolution of about $15\mum$. As a
consequence of the high granularity of the silicon tracker and the
strong and homogeneous magnetic field, a transverse momentum
resolution of about 1.5\%~\cite{Khachatryan:2010pw} is obtained
for the muons in this analysis.  The systematic uncertainty in the
track reconstruction efficiency for charged hadrons is estimated to be
4.0~(2.3)\% in Run~1
(Run~2)~\cite{Khachatryan:2010pw,CMS-DP-2018-050}.
In Run 2, the microstrip detector experienced operational instabilities
during the initial period of the run, resulting in a significant
impact on the trigger efficiency as the level of pileup increased.
The Run 2 data are therefore divided into two separate running periods,
denoted 2016A and 2016B, of roughly equal integrated
luminosity. Separate sets of MC samples are used to describe these
periods.
Residual differences between the MC simulation and the Run~2
data lead to a systematic uncertainty of $\pm0.07\ps$ in the effective
lifetime measurement. This uncertainty is estimated from the variation
in the \Bp\ lifetime measured with the \Bp\ normalization sample in
the two running periods and channels. The maximum difference with respect to the
result in Ref.~\cite{pdg2018} is assigned as the systematic uncertainty.
The uncertainties due to the residual misalignment of the tracker have
negligible impact on the branching fraction measurement. For the
effective lifetime measurement, a systematic uncertainty of 0.02\ps\ is
determined from this source, using the normalization sample.

Muons are measured within $\abs{\eta} < 2.4$ with four muon stations
interspersed among the layers of the steel flux-return plates. Each
station consists of several layers of drift tubes and cathode strip
chambers in the regions $\abs{\eta} < 1.2$ and $0.9 < \abs{\eta} < 2.4$,
respectively. They are complemented by resistive plate chambers (RPC)
covering the range $\abs{\eta} < 1.6$. The muon system does
not contribute to the \pt\ measurement of the muons relevant for this
analysis and is used exclusively for trigger and muon identification
purposes.
Standalone muons are reconstructed from hits in the three muon
subdetectors.
They are subsequently combined with tracks found in
the silicon tracker to form global
muons~\cite{Chatrchyan:2012xi,Sirunyan:2018}. For a global muon, a
standalone muon is linked to a  track by comparing their
parameters after propagation to a common surface at the
innermost muon station of the reconstructed standalone muon track.

\section{Trigger}
\label{s:trigger}
Events with dimuon candidates are selected using a two-tiered trigger
system~\cite{Khachatryan:2016bia}. The first level (L1), composed of
custom hardware processors, uses information from the muon detectors
and calorimeters to select events at a rate of around 100\unit{kHz} within
a time interval of less than 4\mus. The second level, known as the
high-level trigger (HLT), consists of a farm of processors running a
version of the full event reconstruction software optimized for fast
processing, and reduces the event rate to around 1\unit{kHz} before
data storage.

The L1 trigger requires two muon candidates with either no
\pt\ requirement or a loose
requirement of $\pt>3\GeV$, depending on the running
period. However, there is an implicit \pt\ threshold of about
$3.5\GeV$ in the barrel and $2\GeV$ in the endcaps since muons must
reach the muon detectors. In Run~1, no
restriction was imposed on the muon pseudorapidity, while in Run~2
both muons were required to be within $\abs{\eta} < 1.6$. As the
instantaneous luminosity increased over the course of Run~1, the trigger selection was
gradually tightened, an effect that is accounted for in the
simulation of the trigger performance. In Run~2, the trigger conditions were
stable for the entire running period. For both running periods, the offline
analysis selection is more restrictive than the trigger requirements.

At the HLT level, the complete silicon tracker information is available,
providing precise muon momentum information for the dimuon invariant
mass calculation and vertex fit. This allows more stringent
requirements to be placed on the single muon and dimuon \pt,
and permits the calculation of the three-dimensional (3D) distance of
closest approach (\dca) between the two muons.

For the signal sample in Run~1, the HLT required the dimuon invariant
mass to satisfy $4.8 < \mll < 6.0\GeV$. The most stringent HLT
selection additionally required $\ptmu>4\GeV$, $\ptmm>3.9\GeV$ ($5.9\GeV$ for
events with at least one muon with $\abs{\eta} > 1.5$), $\dca <
0.5\cm$, and the probability of the $\chi^2$ per degree of freedom (dof) of
the dimuon vertex fit $\calP(\chidof) > 0.5\%$. For Run~2, the
following requirements were imposed: $\abs{\eta} < 1.4$; $4.5 < \mll <
6.0\GeV$; $\ptmm>4.9\GeV$; $\ptmu>4\GeV\ (3\GeV)$ for the leading
(subleading) muon, where the leading (subleading) muon is the muon
with the higher (lower) \pt; $\dca < 0.5\cm$; and
$\mathcal{P}(\chidof) > 0.5\%$.

For the normalization (\bupsikp) and control (\bzspsiphi) samples, the
data in Run~1 were collected by requiring the following: two muons,
each with $\ptmu> 4\GeV$ and $\abs{\eta}<2.2$, $\ptmm > 6.9\GeV$, $2.9
< \mll <3.3\GeV$, $\dca < 0.5\cm$, and $\calP(\chidof) > 15\%$.  To
reduce the rate of prompt \jpsi\ candidates, two additional
requirements were imposed in the transverse plane: (i) the pointing
angle $\alpha_{xy}$ between the dimuon momentum and the vector from
the beamspot (defined as the average interaction point) to the dimuon
vertex must fulfill $\cos\alpha_{xy} > 0.9$; and (ii) the flight
distance significance $\ell_{xy}/\sigma(\ell_{xy})$ must be larger
than 3, where $\ell_{xy}$ is the 2D distance between the beamspot and
the dimuon vertex and $\sigma(\ell_{xy})$ is its uncertainty. For
Run~2, the only changes were to restrict the muons to $\abs{\eta} <
1.4$ and to loosen the vertex probability requirement to
$\calP(\chidof) > 10\%$. In addition, for Run 2, this trigger path was
prescaled by a factor between 1 and 8, depending on the instantaneous
luminosity.

The trigger efficiencies for the various samples are determined from
the MC simulation.  They are calculated after all muon identification
selection criteria, discussed in Section~\ref{s:muonid}, and
analysis preselection criteria, discussed in
Section~\ref{s:selection}, have been applied.  For the signal events,
the average trigger efficiency is around 70\%\ (up to 75\% in the
central channel and down to 65\% in the forward channel). The trigger efficiency for
the normalization and control samples varies from 75\% in the central
channel to 50\% in the forward channel.  A systematic uncertainty of
3\% in the trigger efficiency ratio between the signal and
normalization samples is estimated from simulation by varying
the selection efficiency between a very loose preselection level to a
level such that only 10\% of the preselected events remain.

The operational instabilities of the microstrip detector during the
2016A running period increased the pileup dependence of the HLT. This
affected the normalization sample more strongly than the signal
sample, because of the requirement for the former sample that the
dimuon vertex be well separated from the
beamspot. The pileup-dependent normalization deficit is corrected in
data with per-event weights that depend on the number of reconstructed
$\Pp\Pp$ interaction vertices and $\ell_{xy}$. The systematic uncertainty
associated with this correction amounts to 6\% for the 2016A data and
5\% for 2016B.

\section{Muon identification}
\label{s:muonid}
For the analysis, it is important to maintain a high muon identification efficiency
while minimizing the probability for charged hadrons to be
misidentified as muons.
Achieving a low hadron-to-muon
misidentification rate is especially important in the search for
\bdmm, where the SM branching fraction is roughly an order of
magnitude below that for \bsmm\ and there are additional contributions
to the background from two-body decays of \PB hadrons.

To achieve this goal, a new muon BDT was trained separately for
the Run~1 and Run~2 data, using the \textsc{tmva}
framework~\cite{Hocker:2007ht}. For both data samples, the starting
point for muon identification is the set of global
muons obtained from the standard CMS muon
reconstruction~\cite{Chatrchyan:2012xi,Sirunyan:2018}. In contrast, in the previous
analysis~\cite{Chatrchyan:2013bka}, the starting point for the BDT
was a sample of so-called tight muons, which forms a subset of the full
global muon sample.

The muon BDT training was performed using simulated events derived from
signal samples for the muons and from background samples for the
misidentified charged hadrons.  Statistically independent MC samples
were used to evaluate the performance of the muon BDT in the optimization
process.  The variables used in the new muon BDT can be grouped into
three categories according to whether they are associated with
measurements from the silicon tracker, the muon system, or the combined
global muon reconstruction.

The variables determined with the silicon tracker are sensitive to
the quality of the muon track measurement and exploit the fact
that tracks from charged hadrons and tracks from particles with a decay-in-flight
often have lower quality. These variables are the track
\chidof, the fraction of valid hits divided by the number of
expected hits, the number of layers containing hits, and the
change in track curvature. The changes in track curvature are
identified using a dedicated kink-finding algorithm, which computes
the difference between the predicted and measured azimuthal angle
$\varphi$  of the track at each layer. The values of the squared
$\varphi$-angle differences, divided by their associated squared
uncertainties, are then summed to obtain a discriminating variable.

The variables associated with measurements in the muon system are the
standalone muon \chidof, the standalone muon
compatibility with the muon hypothesis, and a variable quantifying the
time-of-flight error in the RPC muon subsystem.

The variables related to the global muon reconstruction are
the \chidof\ of the momentum matching between the extrapolated
silicon tracker and standalone muon at the innermost muon layer;
the \chidof\ of the position matching between the extrapolated
silicon tracker and standalone muon at the innermost muon layer; the $\chi^2$
between all silicon tracker hit positions and the global muon
position; the output of the kink-finding algorithm (as described
above, but applied to the global muon trajectory); the probability of
the global muon track \chidof; and the product of the charges, as
determined in the silicon tracker and the muon system.

The variables used in the muon BDT optimization process are chosen
iteratively to provide the best performance with a minimal set of
variables (the size of the MC training samples is limited). The same
variable set is used in the muon BDT training for Runs~1~and~2.

The new muon BDT achieves an average misidentification rate of
$6\times10^{-4}$ and $10^{-3}$ for pions and kaons, respectively, for
both Runs~1~and~2, together with a muon efficiency of about 70
(76)\% for Run~1 (Run~2). Compared to the previous BDT used in
Ref.~\cite{Chatrchyan:2013bka}, the new BDT improves the muon efficiency by about 5\%
(absolute) for the same hadron misidentification rate.  The proton
misidentification rate is approximately $10^{-4}$.

The performance of the muon BDT is validated by comparing its behavior
in simulation with that in data, using event samples in which a
kinematically selected two-body decay provides a source of
independently identified muons or hadrons. For muons, the decay
$\jpsi\to\MM$ is used. Charged hadrons are selected with the decays
$\KS\to\pip\pim$ for pions, $\phi\to\Kp\Km$ for kaons, and $\Lambda\to
\p \pim$ for protons. These samples are used to compare the
distributions of the variables used in the muon BDT in
background-subtracted data and simulation, as well as the
corresponding single-hadron misidentification
probabilities. The distributions of all variables
  used in the muon BDT are found to be consistent between data and
  simulation. After correcting for trigger and reconstruction biases,
  the misidentification probabilities in data and simulation are
  also found to be consistent. This comparison is used to assign a 10\% relative
  uncertainty in the pion and kaon misidentification probabilities,
  which are found to be roughly uniform over the range $5 <\pt <
  20\GeV$. The limited statistical precision of the $\Lambda\to \p
  \pim$ validation sample, with a proton misidentified as a muon, does
  not allow a differential comparison, and a relative systematic
  uncertainty of 60\% is estimated based on the average difference between data and
  simulation for the rate of proton misidentification.

An independent study was performed to measure the misidentification
rate of charged pions and kaons by reconstructing
$\Dstarp\to\Dz(\to\Km\pip)\pi^+_{\mathrm{s}}$ decays, where the slow pion
$\pi^+_{\mathrm{s}}$
allows the unambiguous identification of the charged kaon. This
validation sample provides a set of charged hadrons with \pt\ and
impact parameter values comparable to those relevant for the
\bdmm\ analysis. The limited size of the sample allows only a
comparison of the integrated misidentification probabilities, which
agree within the uncertainties between data and simulation.

To determine the systematic uncertainty in the muon identification
efficiency, the muon BDT discriminator distribution is studied in data and
simulation for muons from \bupsikp\ and \bzspsiphi\ candidates. The
efficiency ratio of the muon BDT discriminator requirement for muons from \bupsikp\ and
\bzspsiphi\ decays between data and MC simulation agrees to better than
3\% in all analysis channels. This value is used as the estimate of the
uncertainty in the relative muon identification efficiency between the
signal and normalization modes.

\section{Selection}
\label{s:selection}
The \bmm\ candidate selection starts with two oppositely charged
global muons.  To retain more muon candidates for the development of the
analysis BDT and the validation of the background estimate, the full
muon BDT discriminator requirement is applied only for the extraction
of the final result. Both muons must have $\pt>4\GeV$ and be matched
to muons that triggered the event. The distance of closest
approach \dca\ between the \PB meson candidate tracks is required to
be less than $0.08\cm$. After constraining the two muon tracks to a
common (secondary) vertex, the invariant mass is required to satisfy $4.8 < \mll < 6.0\GeV$.

The momentum and vertex position of the \PB candidate are used to
select the primary vertex (PV) from which the \PB candidate
originates,
based on the distance of closest approach to each PV of the
extrapolated trajectory of the \PB candidate.
In the following, this PV is referred to as the
$\bbbar$-PV. To avoid a possible bias in the $\bbbar$-PV position,
each PV is refit without the \PB candidate tracks.  In this fit, based
on an adaptive fitting method~\cite{Chatrchyan:2014fea}, a weight from 0 to 1
is assigned to each track. The \PB candidate is rejected if the
average track weight of the $\bbbar$-PV (excluding the \PB candidate
tracks) is smaller than 0.6.

In the offline analysis, many of the variables with the highest
discriminating power are determined in 3D space. The flight-length
significance \fls\ is measured with respect to the
$\bbbar$-PV. For the Run 2 analysis, a correction is applied to
\fl\ and \fle\ to reduce differences between data and simulation.
The \PB candidate pointing angle $\alpha$ is calculated
as the opening angle between the
\PB momentum and the vector from the $\bbbar$-PV to the secondary
vertex. The impact parameter \pvip\ of the \PB candidate, its
uncertainty $\sigma(\pvip)$, and its significance \pvips\ are measured
with respect to the $\bbbar$-PV. The \chidof\ of the secondary vertex
fit is also a powerful discriminant. The decay time is given by the product of the flight length \fl\
and the invariant mass of the \PB candidate,
divided by the magnitude of the \PB candidate momentum.

To reduce background, isolation requirements are placed on the
\PB candidate and the muon tracks.  The background rejection power and signal efficiency of
these requirements is not significantly affected by the increased
pileup in Run~2.  The presence of other PVs, not associated with the
\PB candidate, requires that the isolation variables be
calculated using only tracks that are related to the \PB candidate and
its $\bbbar$-PV. In the following, the track sums
include only those tracks that are associated with the $\bbbar$-PV or
that are not associated with any other PV. The latter class of tracks includes
displaced tracks that are not part of any PV, but come close to the
\PB candidate's secondary vertex according to criteria defined below.
Tracks that are part of the \PB candidate are excluded
from the track sums.

The \PB candidate isolation is calculated as $I = \ptb/(\ptb +
\sum_{\text{trk}}\pt)$. The sum includes all tracks from the
    $\bbbar$-PV with $\pt > 0.9\GeV$ and $\Delta R =
    \sqrt{\smash[b]{(\Delta\eta)^2+(\Delta\varphi)^2}}< 0.7$, where
    $\Delta\eta$ and $\Delta\varphi$ are the differences in pseudorapidity and
    azimuthal angle  between the charged track and the
    \PB candidate momentum. Tracks not associated with any other PV
    are also included in the sum if they have $\dca < 0.05\cm$ with
    respect to the secondary vertex. A similar isolation variable,
    $I_\mu = \ptmu/(\ptmu + \sum_{\text{trk}}\pt)$, is calculated for
    each muon, where the requirements $\pt > 0.5\GeV$, $\Delta R <
    0.5$, and $\dca < 0.1\cm$ are imposed.  The track requirements for
    both isolation criteria are chosen to provide optimal background
    rejection and produce good agreement between the data and MC
    simulation.

A variable \closetrk\ is introduced to specify the number of tracks with $\pt >
0.5\GeV$ that satisfy $\dca < 0.03\cm$ with respect to the \PB candidate vertex. This
variable is used to reduce background from partially reconstructed
decays, \eg, $\Bm\to\rho^0(\to\pip\pim)\mun\nub$, where one pion is
misidentified as a muon. The minimum distance of closest
approach \docatrk\ to the \PB candidate secondary vertex is determined
with the same set of tracks, with a correction
applied for Run 2 to reduce data versus simulation differences.

The final selection relies on an analysis BDT
trained~\cite{Hocker:2007ht} with MC signal events (\bsmm\ and
\bdmm\ decays) and with combinatorial background events taken from a data sideband
in the dimuon invariant mass defined by $5.45 < \mll < 5.9\GeV$. The
analysis BDT selection for the Run~1 data is described in
Ref.~\cite{Chatrchyan:2013bka}. For Run~2, the 2016A and 2016B running
periods require separate treatment, as stated above. To avoid possible
bias, the data sample is randomly split into three subsets such that
the training and validation of the analysis BDT are performed on
samples independent of its application. A preselection eliminates
events with extreme outlier values in the relevant variables and
removes the vast majority of background events where the dimuon vertex
is not well separated from the $\bbbar$-PV. The signal and
normalization sample topologies are kept as similar as possible to
reduce uncertainties in their efficiency ratios,
cf.~Eq.~(\ref{eq:schema}).  The most important preselection
requirements are $\fls>4$, $\flsxy>4$, $\pa< 0.2$, $\pvip < 0.02\cm$,
$\pvips < 4$, $\chidof < 5$, and $\dca < 0.08\cm$. After this
preselection, more than 6000 dimuon background events remain in each
subset. Because of this statistical limitation, the analysis BDT
optimization starts from a set of core variables: \fls, \pa, \pvips,
\docatrk, \chidof, \closetrk, \iso, \isomuone, and \isomutwo. To this
list, optional variables, \dca, \pvip, \fl, \flsxy, \pt, and $\eta$,
are added.  The final analysis BDTs are chosen based on the maximum of
$S/\sqrt{S+B}$, where $S$ is the expected \bsmm\ signal yield in the
mass region $5.3 < \mll < 5.45\GeV$ from simulation and $B$ is the expected
combinatorial background in that mass region, extrapolating from the
data sideband. This approximate figure of merit was used only in the
optimization procedure and not in the procedure used to obtain the
final result. Systematic uncertainties related to the selection are
determined using the normalization and control samples, as described
below.

The reconstruction of the \bupsikp\ normalization sample and the
\bzspsiphi\ ($\phi\to\Kp\Km$) control sample is similar to the
reconstruction of \bmm\ candidates. Two oppositely charged global
muons with $\pt >4\GeV$, $\ptmm > 7\GeV$, and $2.8 < \mll < 3.2\GeV$
are combined with either one or two tracks, assumed to be kaons, with $\pt >0.6\GeV$. The
maximum distance of closest approach ($\maxdoca$)
between all pairs of the \PB candidate tracks is required to satisfy
$\maxdoca < 0.08\cm$. For \bzspsiphi\ candidates, the two kaons must
have an invariant mass $1.01 < \mkk < 1.03\GeV$. All \PB candidates
with an invariant mass $4.8 < m < 6.0\GeV$ are retained for further
analysis.  Since \bupsikp\ and \bzspsiphi\ candidates are analyzed with
the same analysis BDT as the \bmm\ candidates, the two muons from the
\jpsi\ are refit to a common vertex and this fit \chidof\ is used in
the analysis BDT, so as to have the same number of degrees of freedom as in
the signal decay. The determination of the other variables is based
on the complete \PB candidate secondary vertex, also including the
additional kaon(s) in the fit.

\begin{figure}[!tb]
  \centering
{\includegraphics[width=0.492\textwidth]{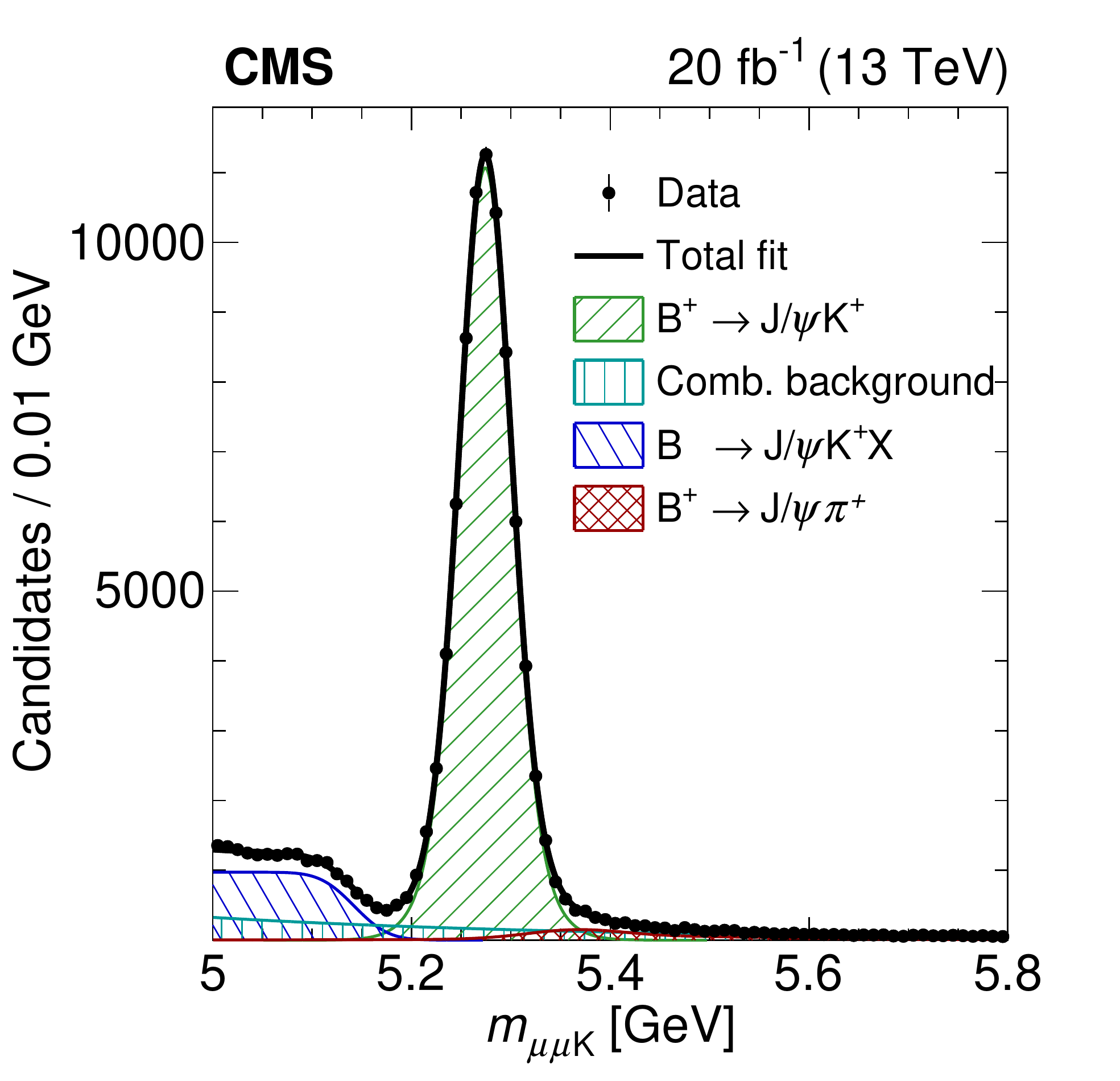}}
{\includegraphics[width=0.492\textwidth]{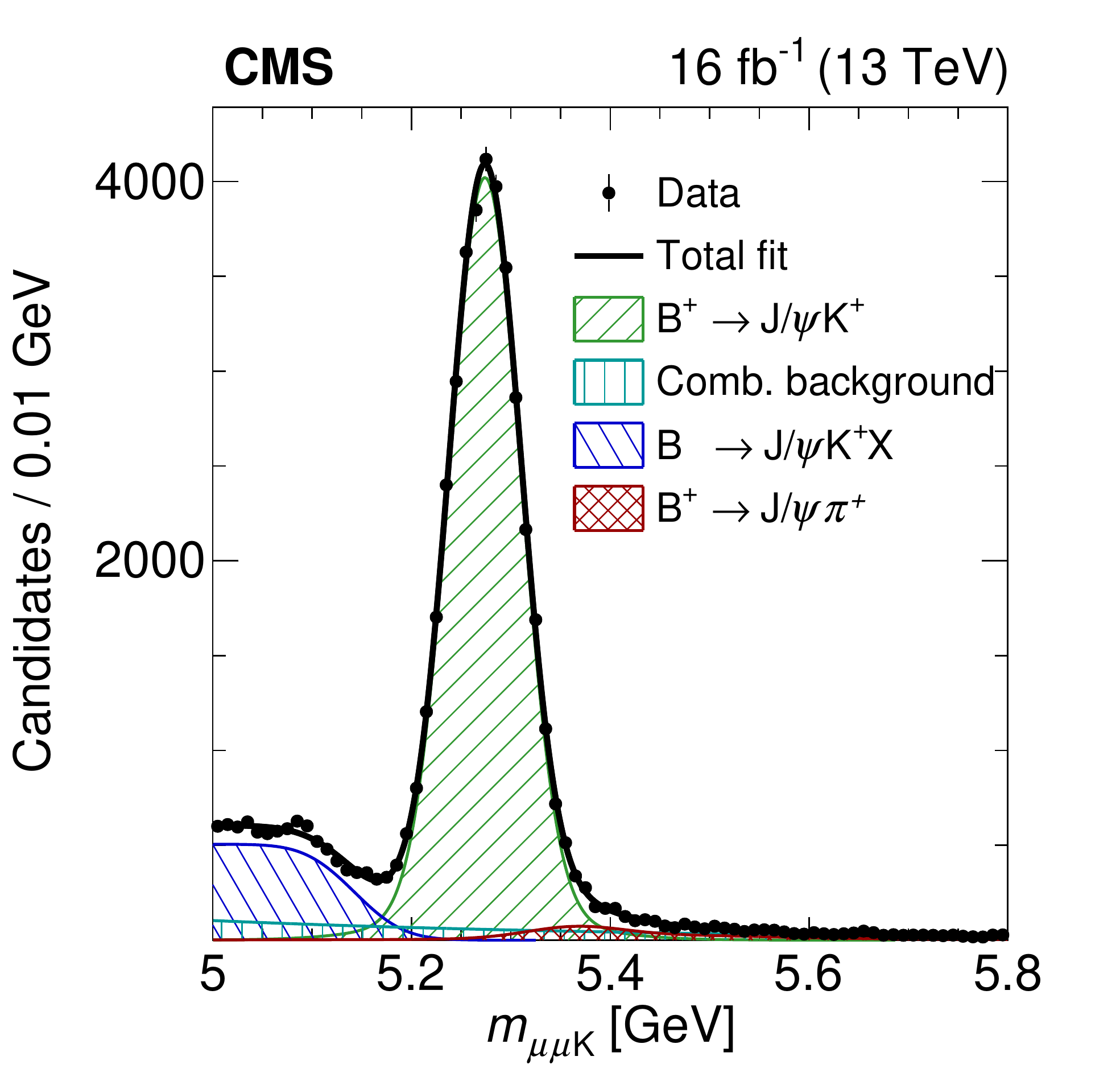}}

  \caption{Invariant mass
    distributions for the $\mu\mu\K$ system used to reconstruct the
    $\bupsikp$ normalization sample. The plot on the left shows the
    2016A central-region channel ($\abs{\etaf} < 0.7$), while the plot
    on the right shows the 2016B forward-region channel ($0.7
    < \abs{\etaf} < 1.4$). The mass resolutions for these channels are
    $30$ and $43\MeV$, respectively. The data are shown by solid black
    circles, the result of the fit is overlaid with the black line, and
    the different components are indicated by the hatched
    regions. } \label{f:normalization}
\end{figure}

The $\B$ candidate yields in the normalization sample are determined
with binned ML fits. Example invariant mass distributions from Run~2
are shown in Fig.~\ref{f:normalization}.  The \bupsikp\ signal
component is modeled by a double-Gaussian function with common
mean. The background is modeled with an exponential function for the
combinatorial component, an error function for the partially
reconstructed background from $\PB\to\jpsi \K X$, and a double-Gaussian function with common
mean for
\bupsipip\ decays.  For this latter component, the integral is
constrained to 4\% of the signal yield~\cite{pdg2018} and the other parameters are
fixed to the expectation from MC simulation. The
  total \bupsikp\ yield used for the determination of $\cbf(\bsmm)$ is
  $N_{\text{obs}}^{\Bp} = \vuse{sum_N_Bu_text} $, where the
  uncertainty combines the statistical and systematic components (see
  Table~\ref{t:bfyields} below for the yields in different running
  periods and channels). The systematic uncertainty in the
\bupsikp\ yield is approximately 4\% and is determined by comparing
the yields when fitting with or without a \jpsi\ mass constraint.

\begin{figure}[!tb]
  \centering
{\includegraphics[width=0.325\textwidth]{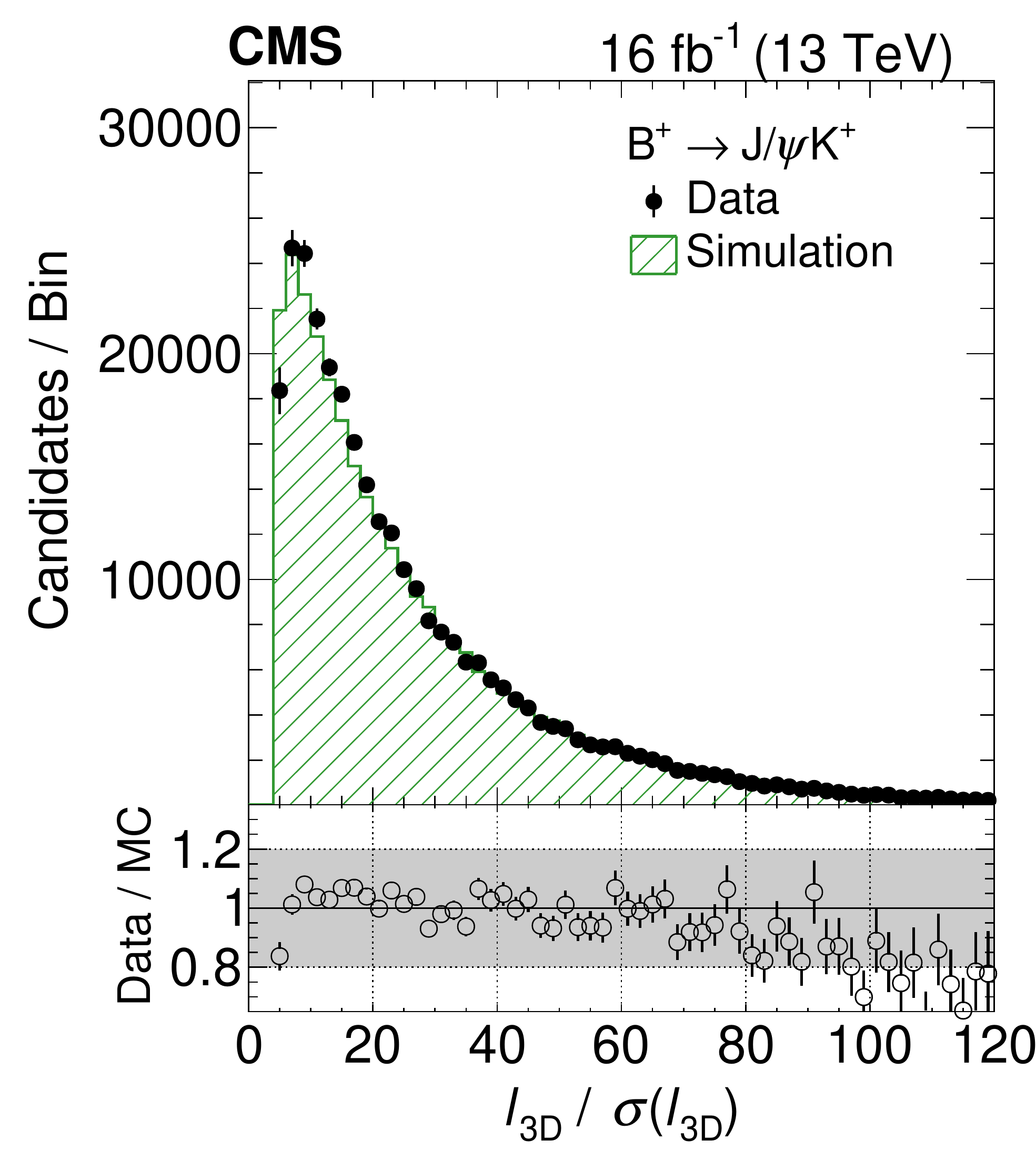}}
{\includegraphics[width=0.325\textwidth]{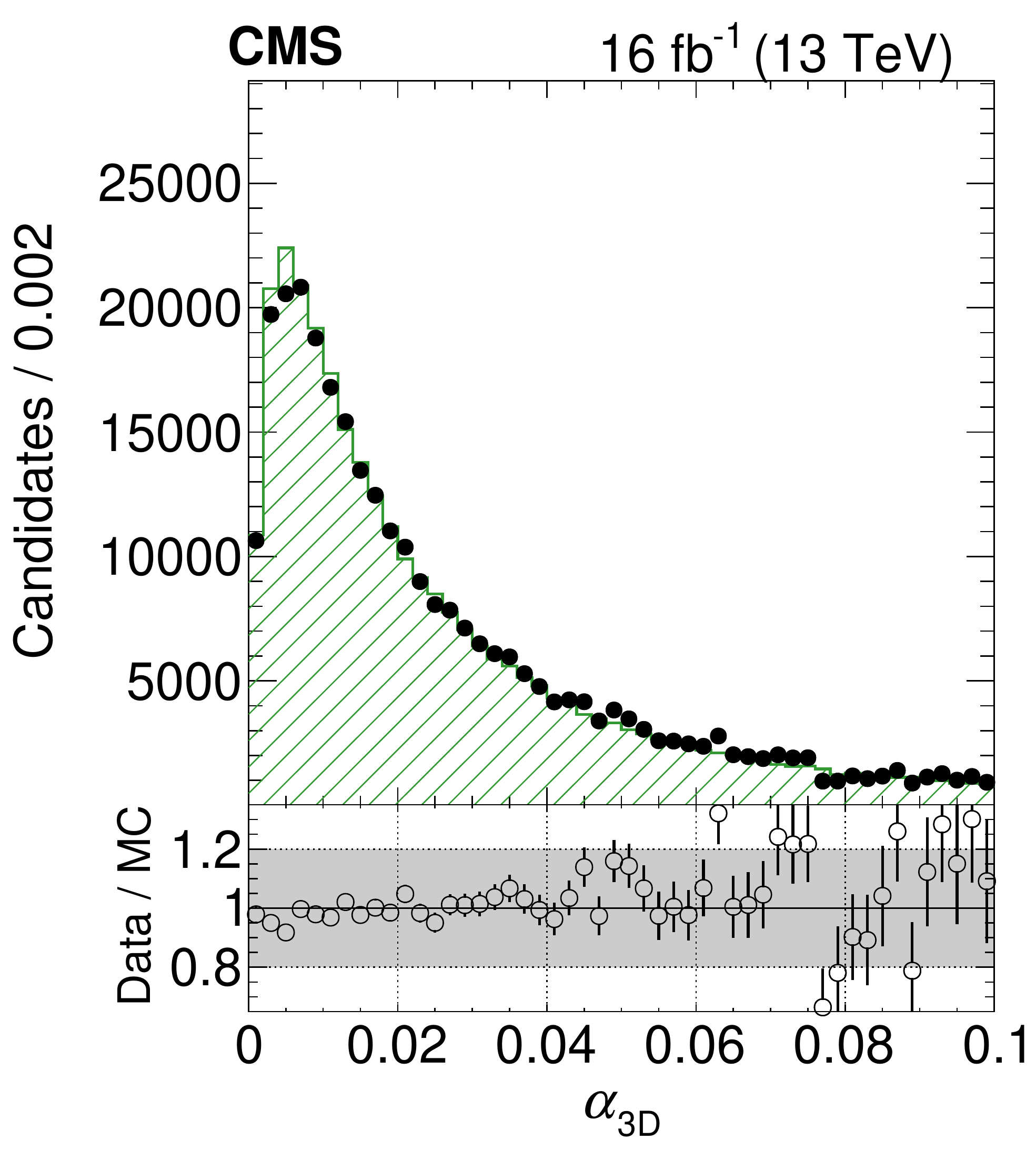}}
{\includegraphics[width=0.325\textwidth]{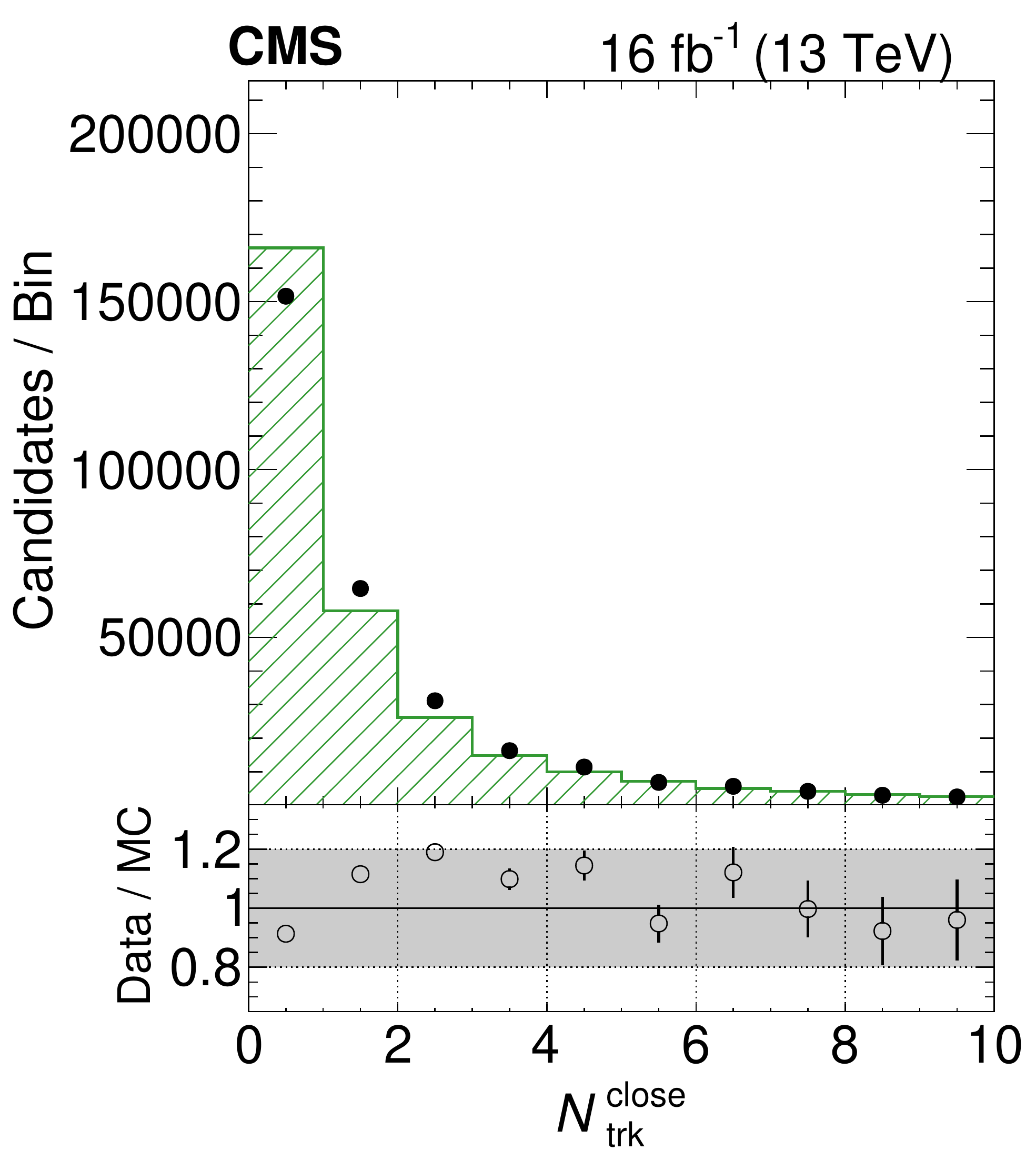}}

  \caption{Comparison of measured and simulated
    \bupsikp\ distributions for the most discriminating analysis BDT
    variables in the central channel for 2016B: the flight length
    significance, the pointing angle, and the number of tracks close
    to the secondary vertex. The events are required to pass the
    preselection for the analysis BDT training. See text for details. The
    background-subtracted data are shown by solid circles and the MC
    simulation by the hatched histogram. The MC histograms are normalized
    to the number of events in the data. The lower panels display the
    ratio of the data to the MC simulation. The band in the ratio plot illustrates a $\pm20\%$ variation.}
  \label{f:datamc1}
\end{figure}

Background-subtracted data distributions are compared to simulation
for both the \bupsikp\ and \bzspsiphi\ candidate event samples. As
examples, Fig.~\ref{f:datamc1} shows the most discriminating variables
used in the analysis BDT: the flight length significance \fls, the
pointing angle~$\alpha$, and the number of close tracks \closetrk. The
distributions are shown for the central channel in the 2016B data
sample after the loose preselection for the analysis BDT training has
been applied. The ratio between the background-subtracted data and the
simulation is shown in the lower plots. The shaded bands in the ratio
plots, included for illustration, demonstrate that the data-simulation
difference can be as large as around 20\%, mostly in peripheral parts
of the distributions (the same comment applies to the analogous bands
shown below in Figs.~\ref{f:datamc2} and~\ref{f:bdtresponse}).  These
residual differences contribute to the systematic uncertainty in the
analysis efficiency ratio between signal and normalization, where
their impact is reduced. In Fig.~\ref{f:datamc2}, distributions of
kinematic variables are displayed: the subleading muon \pt, the cosine of
the muon helicity angle $\theta_{\mun}$ ($\theta_{\mun}$ is the angle
in the \jpsi\ rest frame between the $\mun$ and the $\Kp$ direction),
and the \PB meson candidate proper decay time. The \bupsikp\ analysis
BDT discriminator distributions are plotted in Fig.~\ref{f:bdtresponse}.
To illustrate the discrimination power of the analysis BDT,
Fig.~\ref{f:bdtresponse} also shows the BDT discriminator
distributions from the \mll data sideband and the \bsmm\ signal MC
simulation.

\begin{figure}[!tb]
  \centering
{\includegraphics[width=0.325\textwidth]{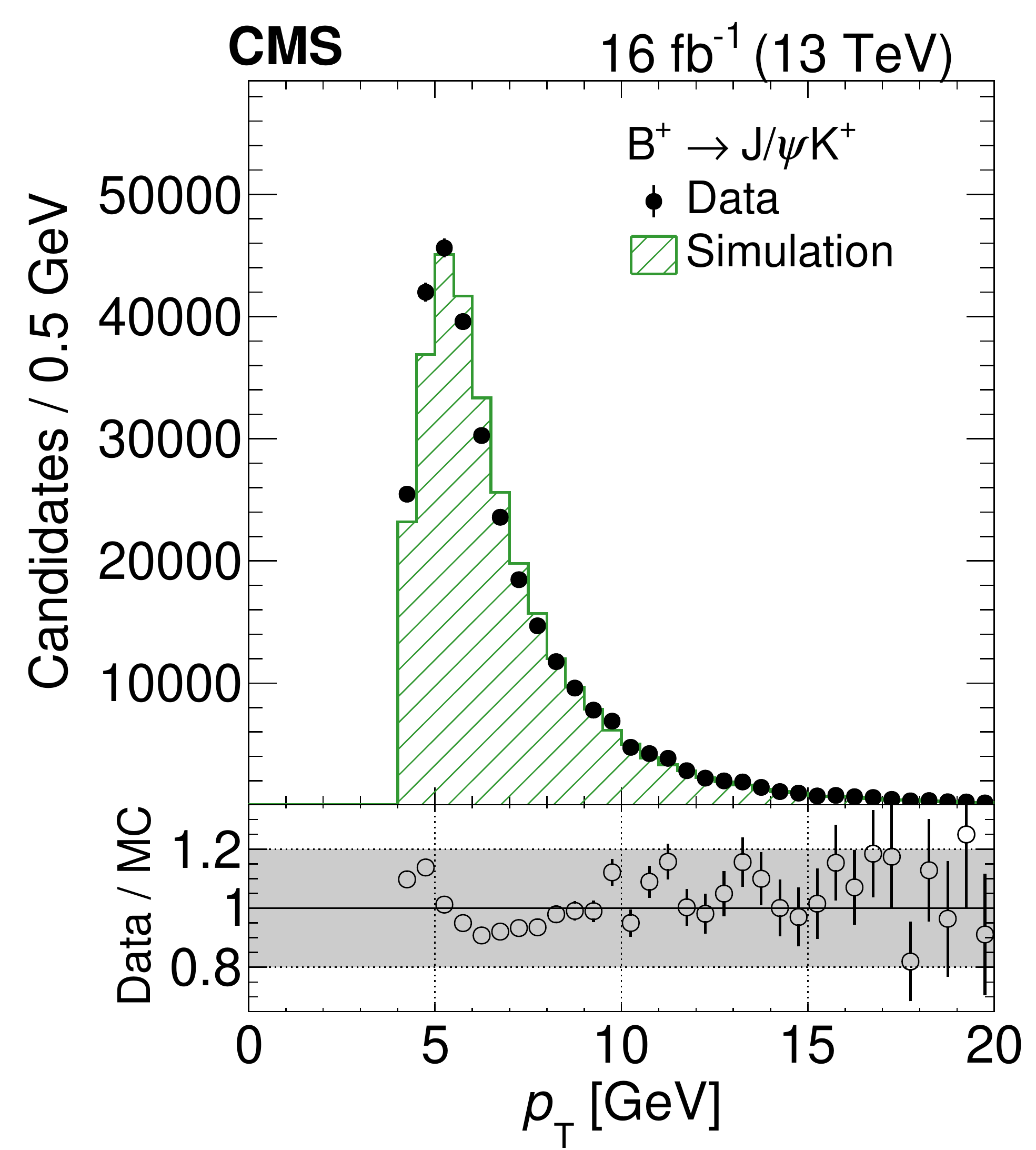}}
{\includegraphics[width=0.325\textwidth]{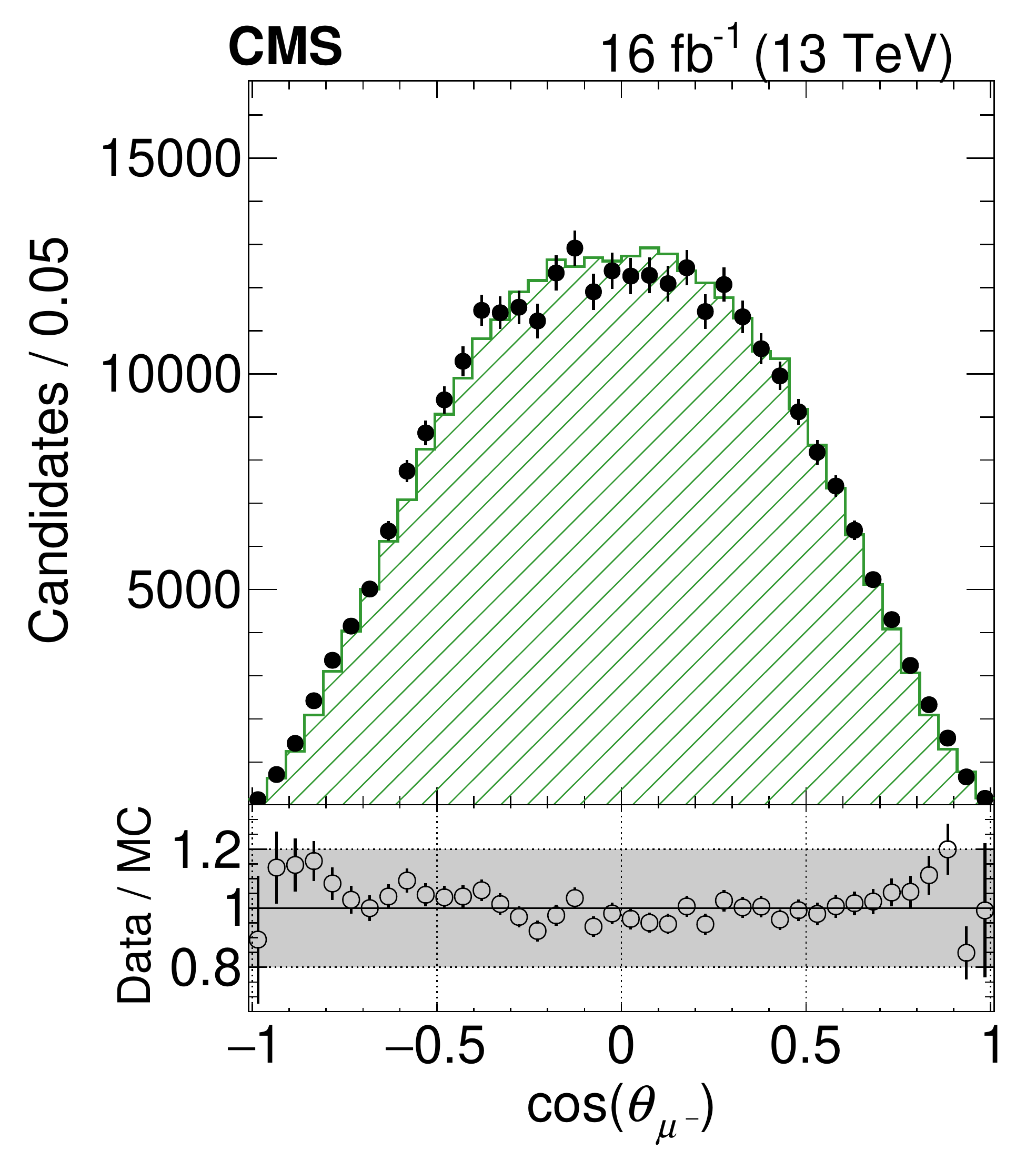}}
{\includegraphics[width=0.325\textwidth]{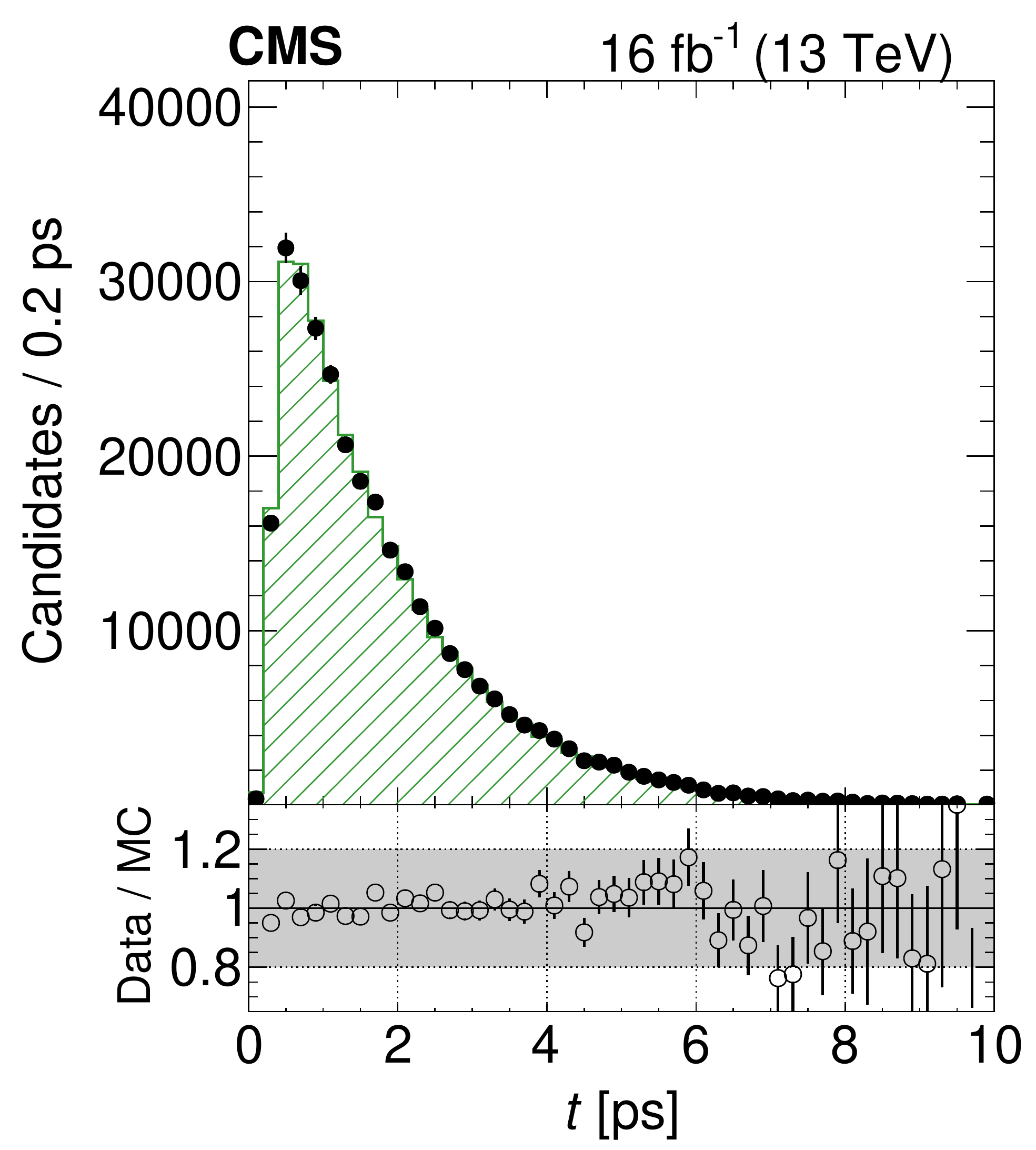}}

  \caption{Comparison of measured and simulated
    \bupsikp\ distributions for kinematic variables
    in the central channel for 2016B: the subleading muon \pt, the
    muon helicity angle, and the \B meson proper decay time. The
    events are required to pass the preselection for the analysis BDT
    training. See text for details. The background-subtracted data are
    shown by solid circles and the MC simulation by the hatched
    histogram. The MC histograms are normalized to the number of
    events in the data. The lower panels display the
    ratio of the data to the MC simulation. The band in the ratio plot illustrates a $\pm20\%$
    variation.}
  \label{f:datamc2}
\end{figure}

The systematic uncertainty in the analysis efficiency (accounting for
losses in reconstruction, identification and selection) and in its ratio
between the signal and normalization modes is estimated with the
double ratio of analysis efficiencies between \bupsikp\ and \bzspsiphi\ decays in
data and MC simulation, based on the distributions in
Fig.~\ref{f:bdtresponse}. The \bzspsiphi\ control sample is used in this
context as a placeholder for the signal because of the statistical
limitations of the data signal events.  Just as the normalization
sample has one additional track compared to the signal sample, the
control sample has one additional track compared to the normalization
sample. Applying the analysis BDT discriminator requirements, the
efficiency ratios are determined between the \bupsikp\ and \bzspsiphi\
samples and subsequently the ratios between data and MC simulation are
calculated for the double ratio. The deviation from unity is taken as
the systematic uncertainty.  In Run~2, it is approximately 5\%, while
in Run~1, it varies between 7 and 10\% depending on the year and
channel.  For the effective lifetime determination, a small systematic
uncertainty of 0.02\ps associated with the selection efficiency is
estimated from a variation of the analysis BDT requirement.

\begin{figure}[!tb]
  \centering

{\includegraphics[width=0.325\textwidth]{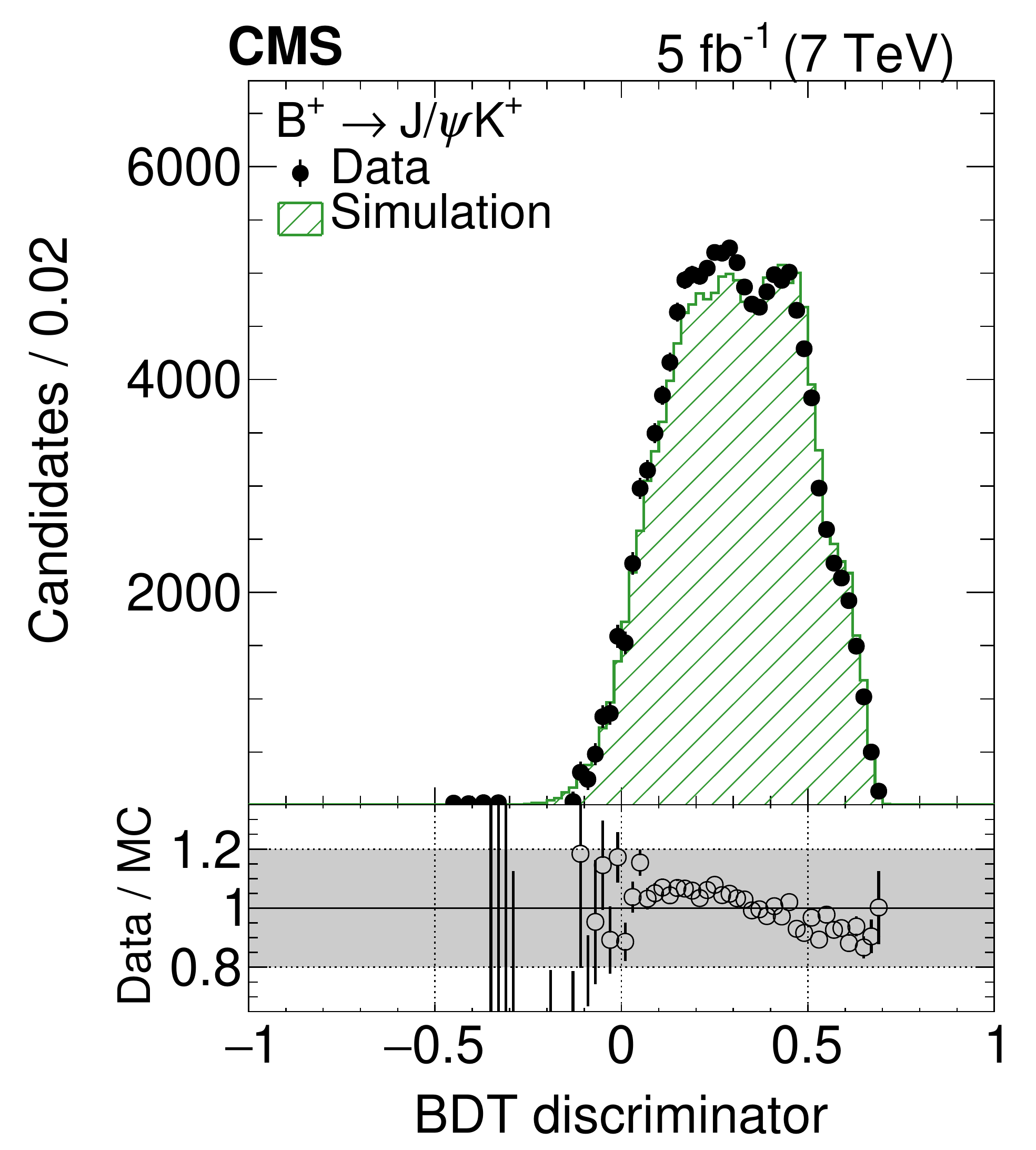}}
{\includegraphics[width=0.325\textwidth]{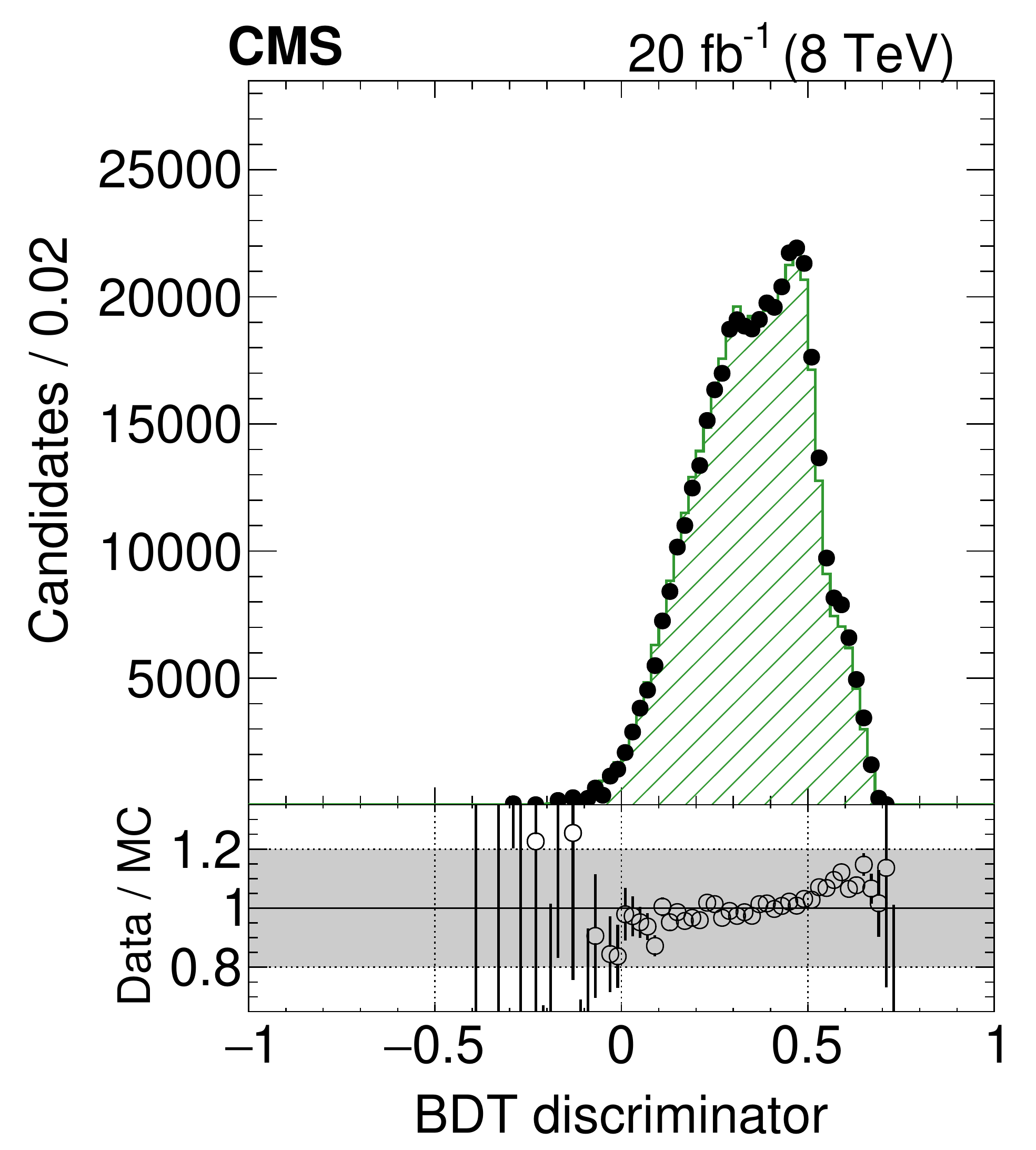}}
{\includegraphics[width=0.325\textwidth]{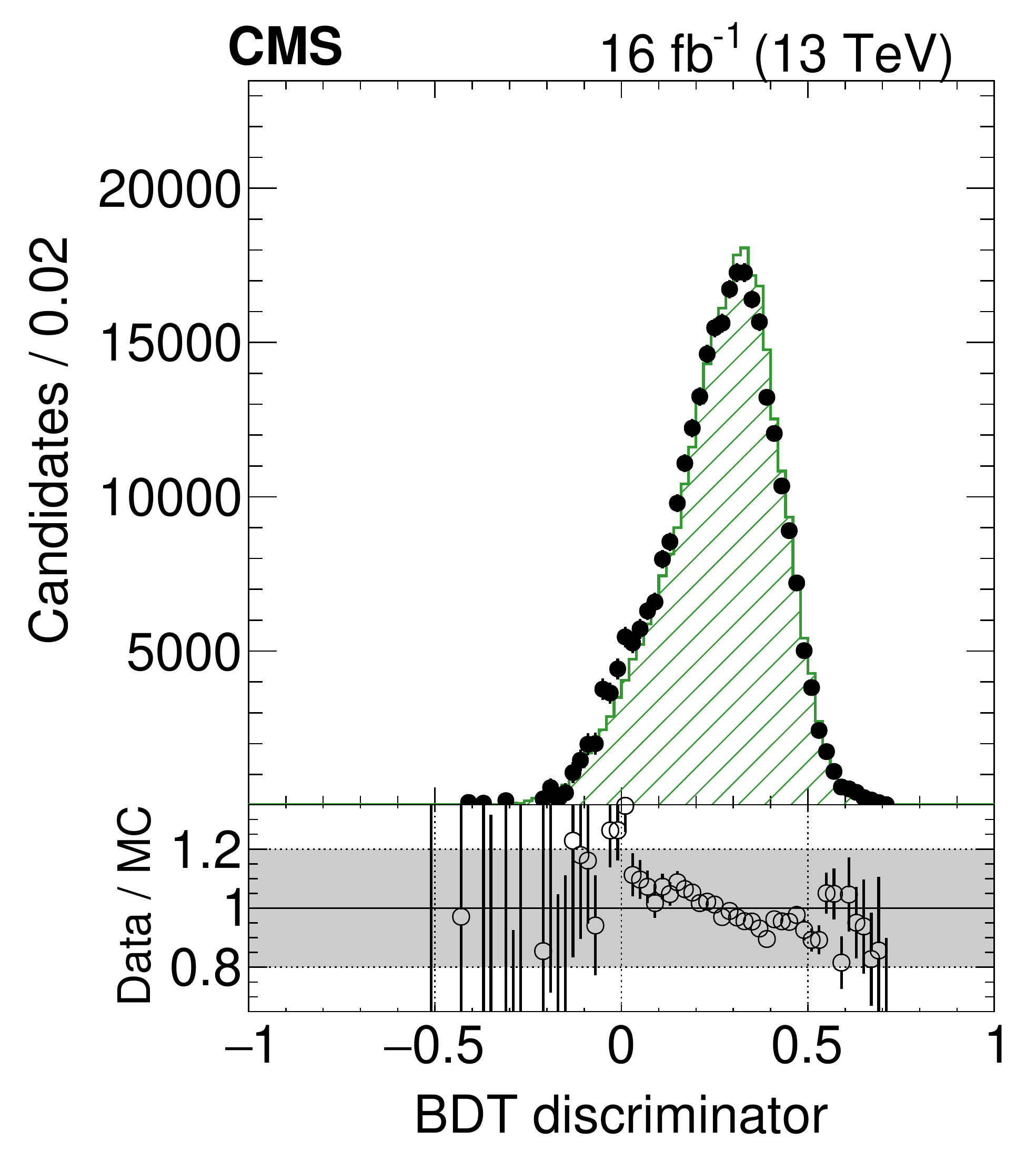}}

{\includegraphics[width=0.325\textwidth]{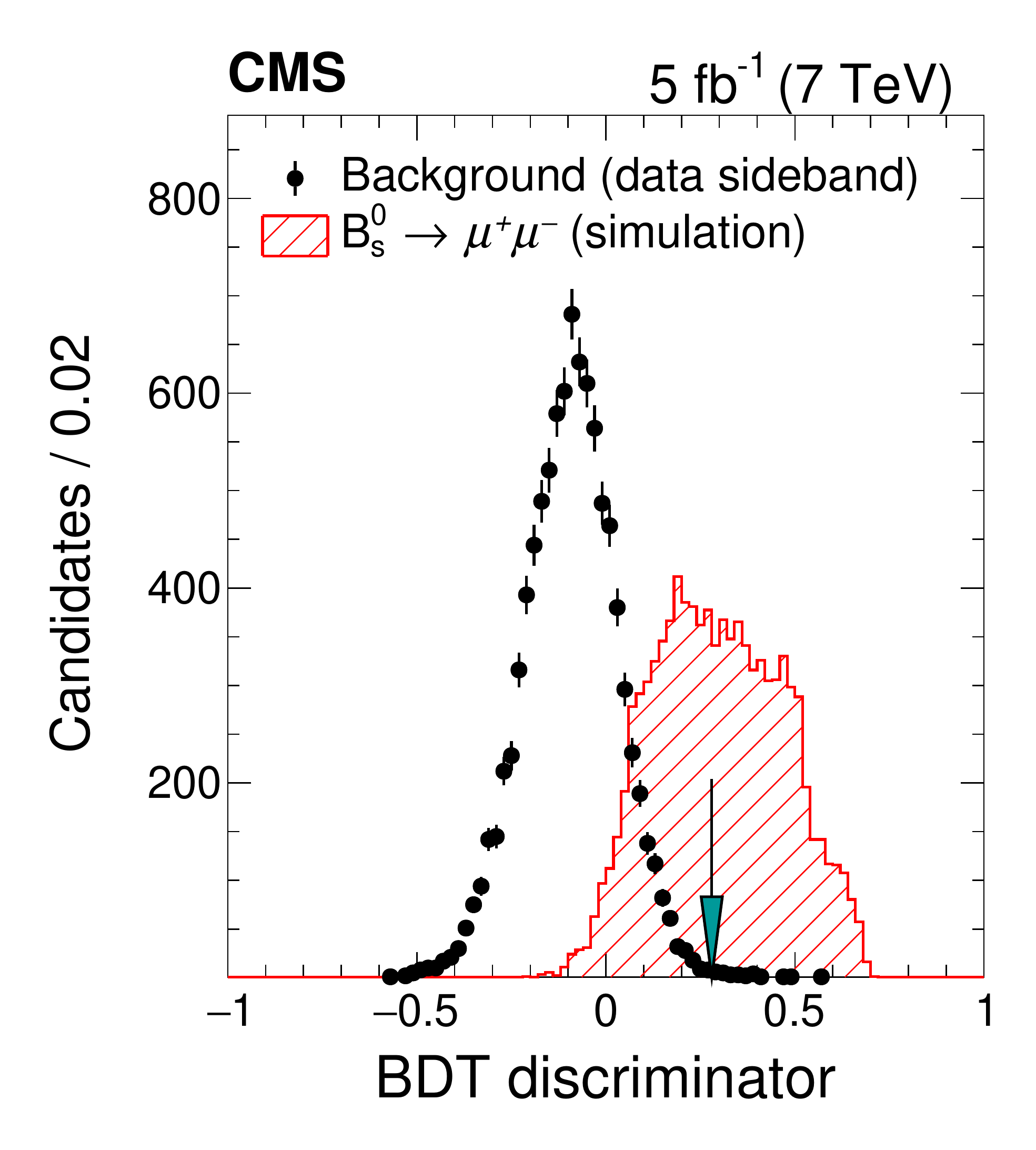}}
{\includegraphics[width=0.325\textwidth]{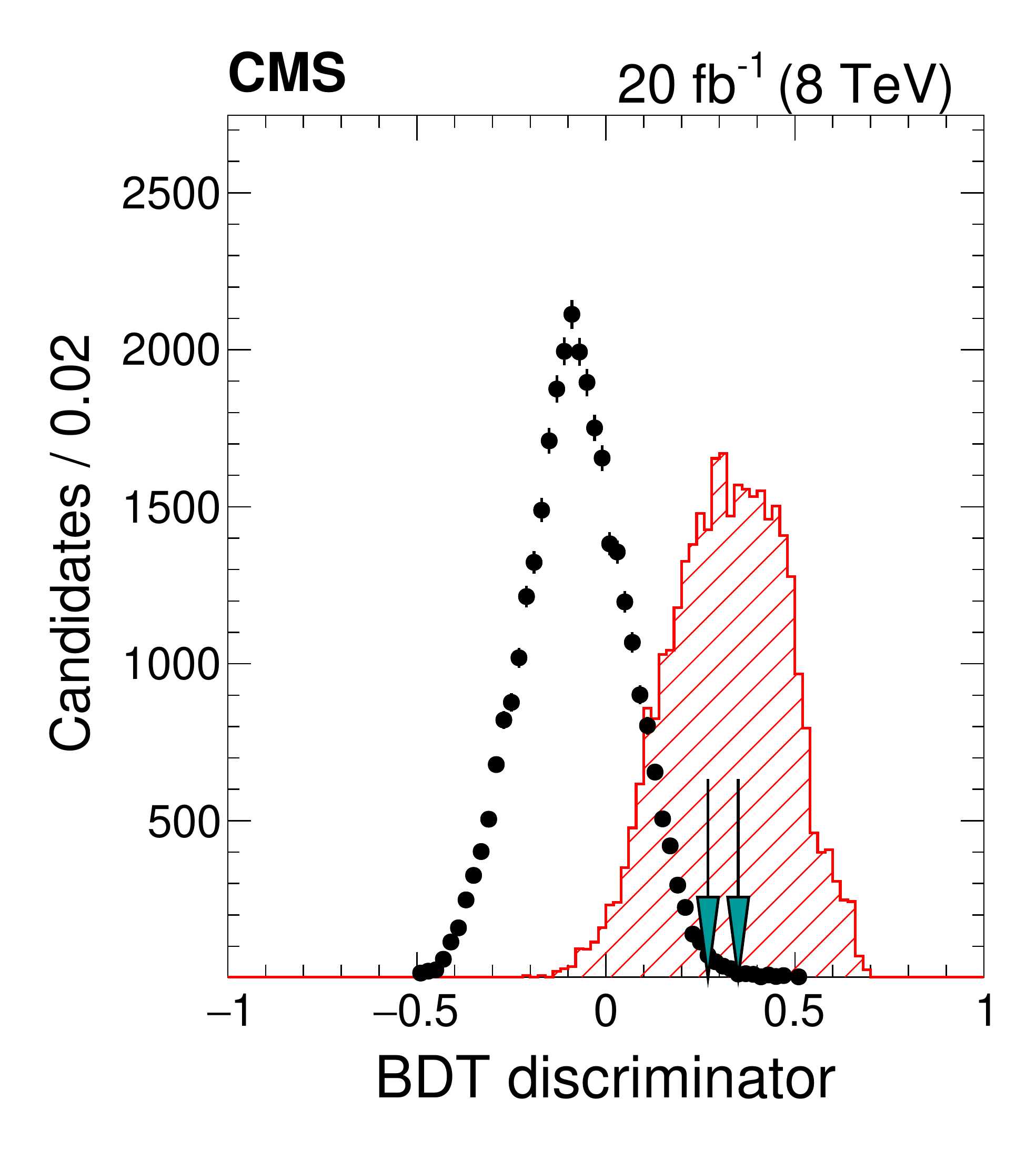}}
{\includegraphics[width=0.325\textwidth]{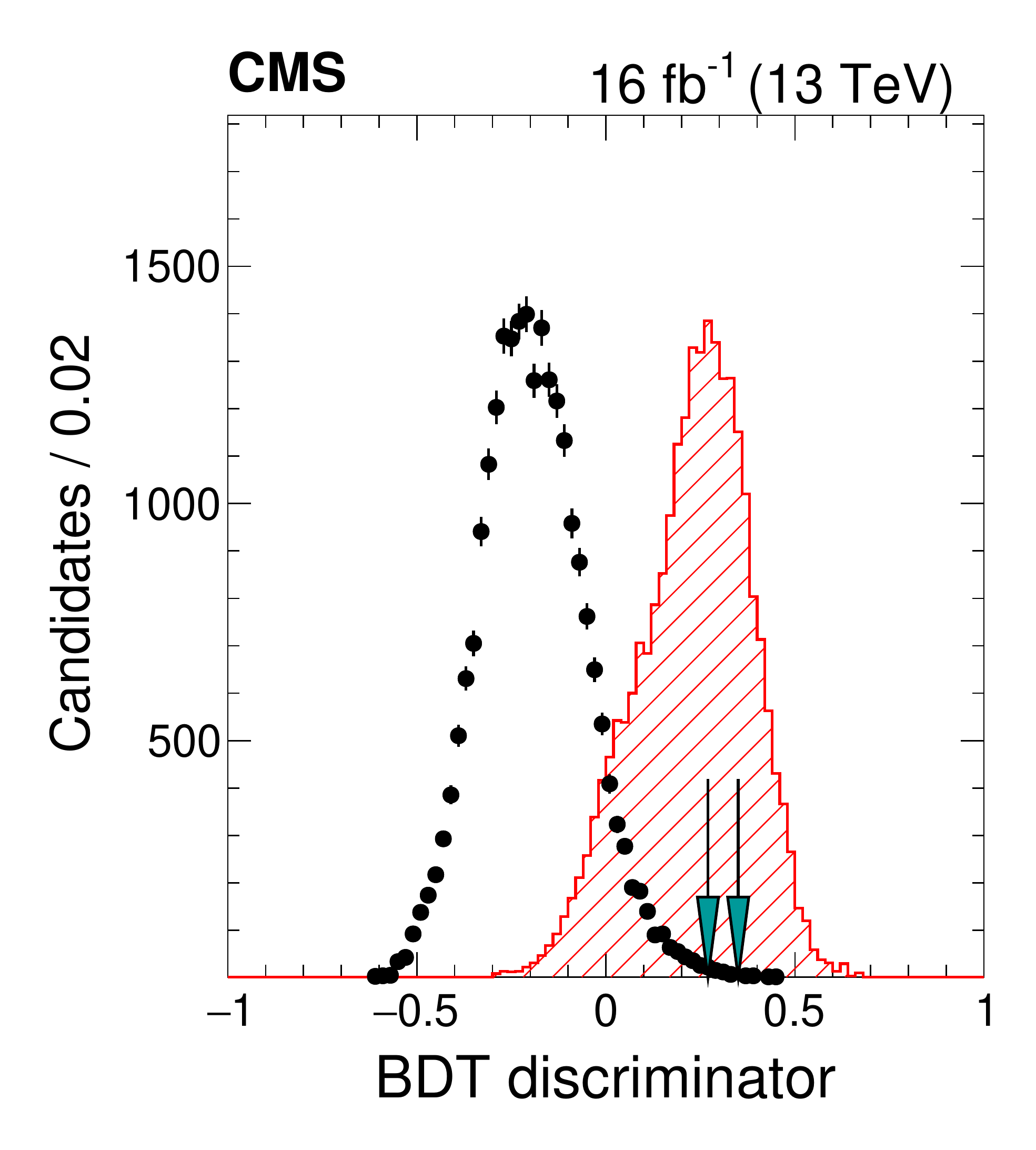}}

  \caption{(Top row) Comparison of the analysis BDT
      discriminator distributions for
      \bupsikp\ in background-subtracted data and MC simulation in the
      central channel for 2011 (left column), 2012 (middle column),
      and 2016B (right column). The lower panels display the
      ratio of the data to the MC simulation. The
      band in the ratio plot illustrates a $\pm20\%$ variation. (Bottom row) Illustration of the
      analysis BDT discriminator distribution in dimuon background
      data from the $5.45 < \mll < 5.9\GeV$ sideband and
      \bsmm\ signal MC simulation. The distributions correspond to the
      full preselection and are normalized to the same number of
      entries. The solid markers show the data and the hatched
      histogram the MC simulation.  The arrows show the BDT
      discriminator boundaries provided in Table~\ref{t:bfbdt}.
  }  \label{f:bdtresponse}
\end{figure}

The signal selection efficiency depends on the
proper decay time because of selection requirements in the trigger and
offline analysis. In the HLT, the increased instantaneous luminosity in Run~2
led to selection requirements on the maximum impact parameter of
tracks relative to the interaction point. These requirements gradually
reduce the efficiency for proper decay times larger than $6\ps$. The analysis
selection requirements on isolation and on the separation of the \PB candidate
secondary vertex from the $\bbbar$-PV strongly reduce the efficiency
for proper decay times below $1\ps$. The events in the signal MC simulation
are reweighted to correspond to the SM expectation of
$\tau_{\mup\mun}^{\text{SM}} = \tau_{\BsH} = 1.615\ps$~\cite{pdg2018}. A
priori, the effective lifetime is unknown, and we therefore use $\Delta
\equiv [\varepsilon_{\text{tot}}(\tau_{\BsH}) - \varepsilon_{\text{tot}}(\tau_{\BsL})]/\sqrt{12}$,
where $\tau_{\BsL}= 1.415\ps$~\cite{pdg2018}, as the uncertainty
associated with the effective lifetime uncertainty. It amounts to
1--3\% in $\cbf(\bsmm)$, depending on the channel and running period.
For the \Bz\ case, where the lifetime difference is much smaller, the average lifetime is used.

The systematic uncertainty due to differences between data and
simulation for the production mechanism mixture is estimated as
follows. In events with a \bupsikp\ or \bzspsiphi\ candidate and a
third muon $\mu_3$,
presumed to originate from the decay of the other \PQb\ hadron in the event,
the variable $\Delta R(B, \mu_3)$ provides
discrimination  between gluon splitting on the one hand and gluon
fusion plus flavor excitation on the other. Templates from MC
simulation are fit to the data distribution and used to determine the
relative production mechanism fractions in data. Reweighting the
fractions in the MC simulation to correspond to the fractions
determined in data provides an estimate of 3\% for the systematic
uncertainty in the efficiency ratio.

To cross-check whether a significant variation in \fsfu\ is observed
with respect to the \PB candidate kinematic variables $\eta$ (up to
$\abs{\eta} < 2.2$) and \pt\ (from 10 to 100\GeV),
efficiency-corrected ratios of \bzspsiphi\ and \bupsikp\ yields are
studied. In addition, a sample of \bdpsikstarz\ (with
$\kstarz\to\Kp\pim$) decays is reconstructed to perform the analogous
cross-check for \fsfd, closely following the procedure discussed in
Ref.~\cite{Sirunyan:2017nbv}. No significant \pt\ or $\eta$ dependence
is observed in either case.

A complementary approach for the study of the systematic uncertainty
in the selection efficiency is to consider the standard deviation of
the effective branching fraction $\cbf(\bzspsiphi)$
over all running periods and channels. For this purpose, the control
sample yield is substituted for the signal yield in
Eq.~(\ref{eq:schema}). The branching fraction $\cbf(\bzspsiphi)$ is
affected by systematic uncertainties in the analysis BDT selection
efficiency and yield determination, the kaon tracking efficiency, and
possible differences between the isolation of \Bp\ and \Bs\ mesons due
to their different fragmentation. A standard deviation of 4\% is
observed, substantially smaller than the combination of the above
systematic uncertainty contributions. We conclude that the systematic
uncertainties are not underestimated.

\section{Branching fraction measurement}
\label{s:bf}
The branching fractions $\cbf(\bsmm)$ and $\cbf(\bdmm)$ are determined
with a 3D extended UML fit to the dimuon invariant mass $\mll$ distribution,
the relative mass resolution $\sigma(\mll)/\mll$, and the binary
distribution of the dimuon pairing configuration $\calc$, where $\calc
= +1\ (-1)$ when the two muons bend towards (away from) each other in
the magnetic field.

The fit model contains six components: the distributions of the
signals \bsmm\ and \bdmm, the peaking background \bhh, the rare
semileptonic background \bhm, the rare dimuon background \bhmm, and
the combinatorial background. The expected yields of the rare
background components are normalized to
the \bupsikp\ yield, corrected for the relative efficiency ratios and
the respective production ratios. The PDF for each component $i$
(where $1 \le i \le 6$) is
\begin{linenomath}
  \begin{equation}
    P_i(\mll,\sigma(\mll),\calc) = P_i(\mll; \sigma(\mll))\, P_i(\sigma(\mll)/\mll)\, P_i(\calc),
  \end{equation}
\end{linenomath}
where the $P_i$ terms are the PDFs for the indicated variable.

The binary \calc\ distribution is used in evaluating the possible
underestimation of the \bhh\ background.  The
misidentification probabilities for \h\ and \hprime\ are assumed
to be independent.  However, this assumption is not
necessarily correct if the two tracks overlap in the detector.
A possible residual enhancement of the double-hadron
misidentification probability exists relative to the
assumption that this probability factorizes into the product
of two separately measured misidentification probabilities.
Potential bias from this source can be reduced by
requiring that both muon candidates satisfy very strict
quality criteria or by demanding that the tracks be spatially
separated.  The effect is different for two muon candidates
that bend towards or away from each other, and thus the \calc\
distribution is introduced.  A scale factor is used in the fit
to account for the change in the hadron misidentification
rate if the hadron bends towards another muon candidate.

The signal PDFs are based on a Crystal Ball
function~\cite{bib-crystalball} for the invariant mass and a
nonparametric kernel estimator~\cite{Cranmer:2000du} with
a superposition of Gaussian kernels for the relative mass resolution. The width $\sigma_{\text{CB}}$ of
the Crystal Ball function is a conditional parameter with a linear
dependence on the dimuon mass resolution $\sigma_{\text{CB}} =
\kappa\, \sigma(\mll)$.
All parameters, except for the normalizations,
are fixed to values obtained from fits to the signal MC
distributions. The dimuon mass scale is studied with
$\jpsi\to\mup\mun$ and $\ones\to\mup\mun$ decays, interpolated to
$m_{\Bs}$. In Run~1, the MC PDF is shifted by $-6.0\ (-7.0)\MeV$ at
the \Bs\ mass for the central (forward) channel, while in Run~2 the
shift is $-4.4\ (-3.1)\MeV$.  The difference in the mass resolution
between data and MC simulation is studied with \bupsikp\ events and is
found to be 5\%. Since correcting for this difference changes the measured
branching fractions by only around 0.2\%, no associated systematic uncertainty is
assigned.

The peaking background is constructed from the sum of all $\bhh$ decay
modes, weighted by the branching fraction, the product of the
single-hadron misidentification probabilities, and the respective
production ratios. We assume that the selection efficiency for the
peaking background equals the signal selection efficiency. The trigger
efficiency is taken as half the signal efficiency, with a 100\% relative
uncertainty because of the limited size of the MC event samples where
both hadrons are misidentified as muons. The invariant mass PDF is
modeled with the sum of a Gaussian and a Crystal Ball function with a
common mean value. The relative mass resolution is modeled with a
kernel estimator as for the signal PDF. The width of the mass
distribution is independent of the per-candidate mass resolution and
is fixed to the distribution obtained from the weighted sum of
background sources in the MC simulation.

The shapes of the mass and relative mass resolution distributions for
the rare semileptonic background are obtained by adding the MC
expectations for $\Bz\to\pim\mup\nu$, $\Bs\to\Km\mup\nu$, and
$\PGLb \to \p \mun \nub$ decays, with a weighting as for the peaking background
(with the exception of the trigger efficiency, which is taken to be
equal to the signal efficiency). Both the mass and relative mass
resolution distributions are modeled with kernel estimators. Rare dimuon background estimates are
based on the decays $\Bz\to\piz\mup\mun$ and $\Bm\to\pim\mup\mun$,
with the mass and relative mass resolution PDFs modeled with
nonparametric kernel estimators.  To account for missing contributions
and efficiency differences with respect to the signal, these background components
(\bhm and \bhmm) are scaled with a common
factor such that their sum, when added to the combinatorial background,
which is extrapolated from the sideband, matches the event yield of data events
in the mass region $4.9 < \mll < 5.2\GeV$.

The normalizations of all of the rare background components are constrained
within the combined uncertainties of the branching fractions,
misidentification probabilities, and efficiencies. For the peaking
background, the combined relative uncertainty is about 100\%,
while the rare semileptonic and dimuonic background components have
relative uncertainties of order 15\%.

The combinatorial background invariant mass distribution is modeled by
a nonnegative Bernstein polynomial of the first degree, whose
parameters (both slope and normalization) are determined in the fit.
The fit result changes by 2.3\% (0.6\%) for \bsmm\ (\bdmm) when using
an exponential function instead.  This is included as a systematic
uncertainty. The relative mass resolution is modeled by a kernel
function, determined from the events in the mass data sideband.

In the UML fit, the parameters of interest are $\cbf(\bsmm)$ and
$\cbf(\bdmm)$. The nuisance parameters are profiled, subject to
constraints. Gaussian constraints are used for the uncertainties in
$\cbf(\bupsikp)$,
the ratio of efficiencies
$\varepsilon_{\text{tot}}^{\Bp}/\varepsilon_{\text{tot}}$, and the
yields of the normalization sample. For the smaller yields of \bhm,
\bhmm, and \bhh\ decays, log-normal priors are used as
constraints. Prior to analyzing the events in the signal region,
extensive tests were performed with
pseudo-experiments to assess the sensitivity of the fitting procedure
as well as its robustness and accuracy.

As mentioned in the Introduction, the data are divided into two
channels (central or forward) depending on the pseudorapidity of the
most forward muon $\abs{\etaf}$, and into data collection running periods (2011,
2012, 2016A, and 2016B). The separation between the central and
forward channels differs between Run~1 and Run~2 because of
limitations imposed by the larger trigger rates in Run~2. In Run~1,
the central (forward) channel covers $\abs{\etaf} < 1.4$\ ($1.4
< \abs{\etaf} < 2.1$) and in Run~2, $\abs{\etaf} < 0.7$\ ($0.7
< \abs{\etaf} < 1.4$). In total, there are eight channels: central and
forward in four data-taking running periods. To maximize the sensitivity,
channels with sufficient statistical precision are divided into
mutually exclusive categories, a low- and a high-range category in the
analysis BDT discriminator. The low-range category extends from the
boundary given in the first row in Table~\ref{t:bfbdt} to the boundary
in the second row. The high range extends from the boundary in the
second row to $+1$. These analysis BDT discriminator requirements are determined by maximizing the
expected sensitivity using the full UML fit framework.  The boundaries
differ between data-taking running periods because the distributions in the
analysis BDT discriminator value differ. In summary, the UML fit is
performed simultaneously in 14 categories.

\begin{table}[!tb]
  \centering
  \topcaption{Analysis BDT discriminator boundaries per category,
    channel, and running period for the branching fraction determination (2011
    has only one category because of the small sample size). Examples of the
    requirements for the central channels are illustrated in Fig.~\ref{f:bdtresponse} (bottom row).}
  \label{t:bfbdt}
  \begin{tabular}{lcccccccc}
    \hline
    &\multicolumn{2}{c}{2011}&\multicolumn{2}{c}{2012}&\multicolumn{2}{c}{2016A}&\multicolumn{2}{c}{2016B}\\
    &Central&Forward&Central&Forward&Central&Forward&Central&Forward\\
    \hline
    Low
    &\NA  &\NA
    &0.27 &0.23
    &0.19 &0.19
    &0.18 &0.23
    \\
    High
    &0.28 &0.21
    &0.35 &0.32
    &0.30 &0.30
    &0.31 &0.38
    \\
    \hline
  \end{tabular}
\end{table}

\begin{figure}[!tb]
  \centering
{\includegraphics[width=0.492\textwidth]{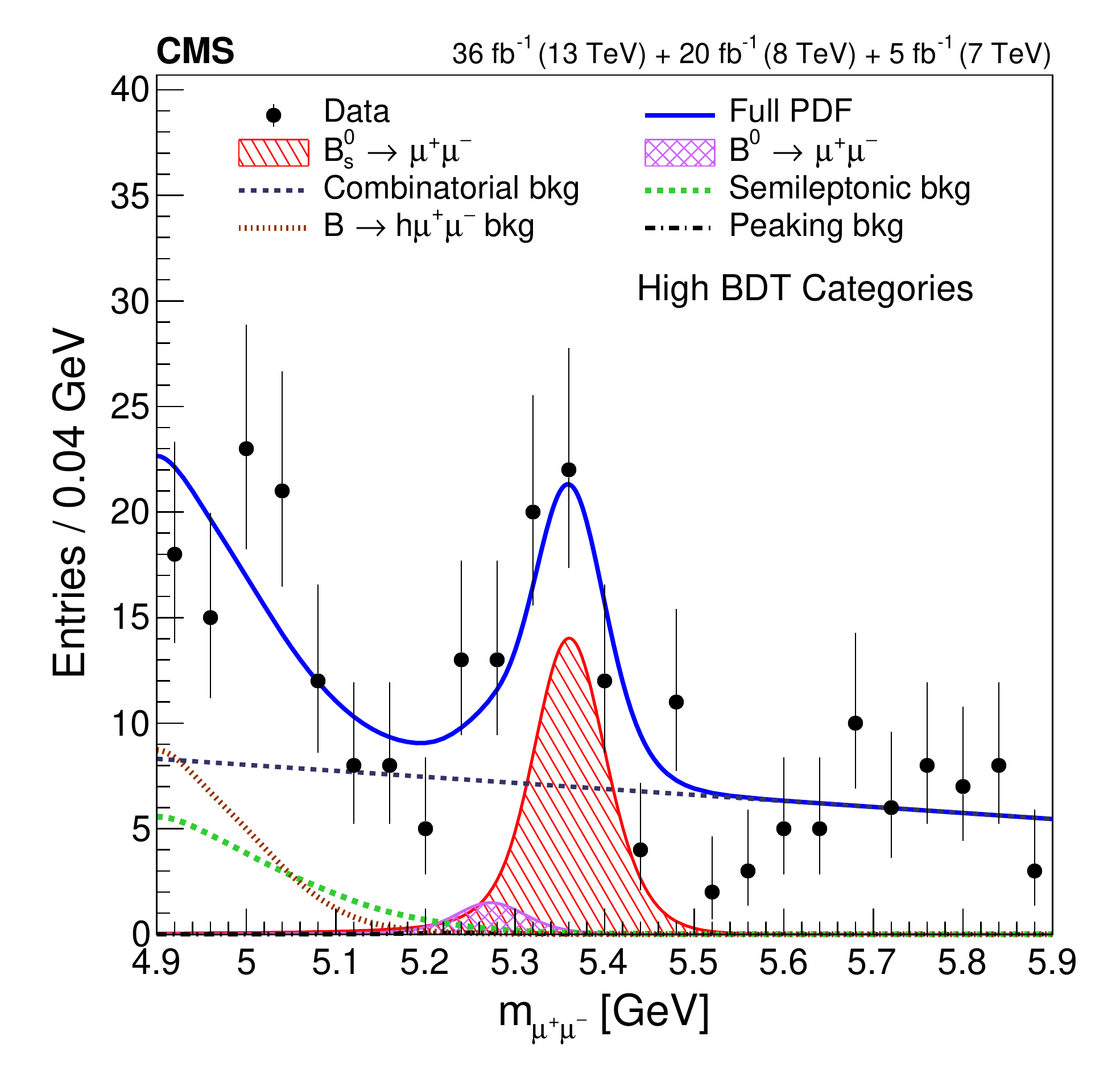}}
{\includegraphics[width=0.492\textwidth]{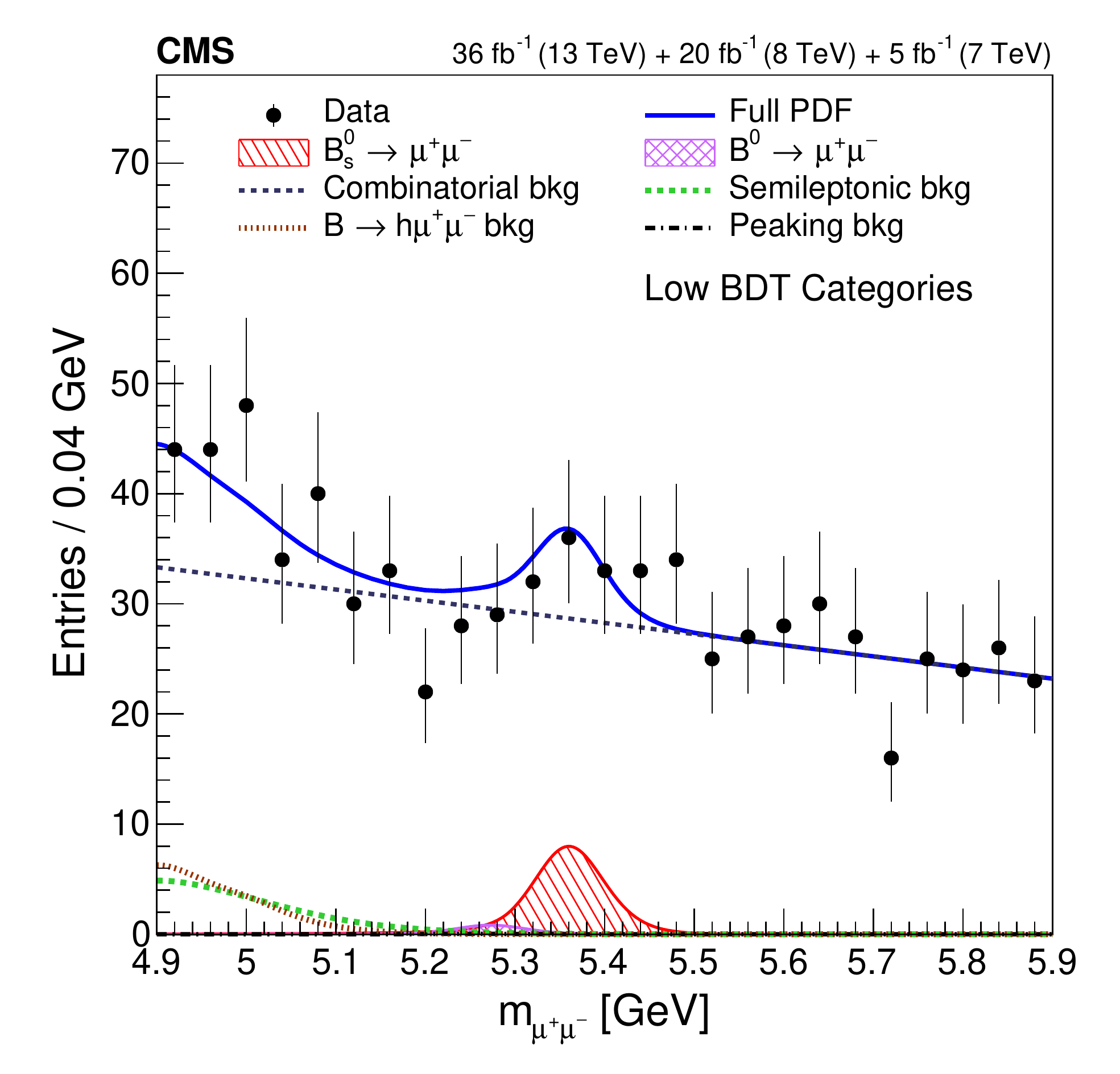}}

  \caption{Invariant mass distributions with the fit projection
    overlays for the branching fraction results. The left (right) plot
    shows the combined results from the high- (low-)range analysis BDT categories
    defined in Table~\ref{t:bfbdt}. The total fit is shown by the
    solid line and the different background components by the broken
    lines. The signal components are shown by the hatched distributions.
  } \label{f:bfmass}
\end{figure}

Figure~\ref{f:bfmass} (left) shows the mass distribution for the
combined high-range analysis BDT categories, with the fit results
overlaid. The \bsmm\ signal contribution is clearly visible.  The
corresponding distribution for the combined low-range BDT categories
is shown in Fig.~\ref{f:bfmass}~(right). The result of the fit to the
data in the 14 categories defined in Table~\ref{t:bfbdt} is
\begin{linenomath}
  \begin{equation}
    \cbf(\bsmm) = \resObsBFBsmm,
  \end{equation}
\end{linenomath}
where the experimental uncertainty combines the dominant statistical
and systematical terms. The second uncertainty is due to the
uncertainty in \fsfu. The systematic uncertainties are summarized in
Table~\ref{t:sys}.  The event yields of the fit components, the
average \pt\ of the \bsmm\ signal, and the efficiency ratios are given
in Table~\ref{t:bfyields}. Summing over all categories, we observe
a total \bsmm\ yield of $\vuse{sum_N_Bs} $
candidates. The observed (expected)
significance, determined using Wilks' theorem~\cite{Wilks:1938dza}, is
$\sigObsBFBsmm$\ ($\sigExpBFBsmm$) standard deviations. The average
$\pt$ of all \bsmm\ signal candidates is 17.2\GeV. The fit also provides the result $\cbf(\bdmm)
= \resObsBFBdmm$. The observed (expected) significance of this result
is $\sigObsBFBdmm$ ($\sigExpBFBdmm$) standard deviations based on
Wilks' theorem, treating $\cbf(\bsmm)$ as a
nuisance parameter. With the Feldman-Cousins
approach~\cite{Feldman:1997qc}, an observed significance of $1.0$ standard
deviation is obtained.

\begin{figure}[!tb]
  \centering
{\includegraphics[width=0.492\textwidth]{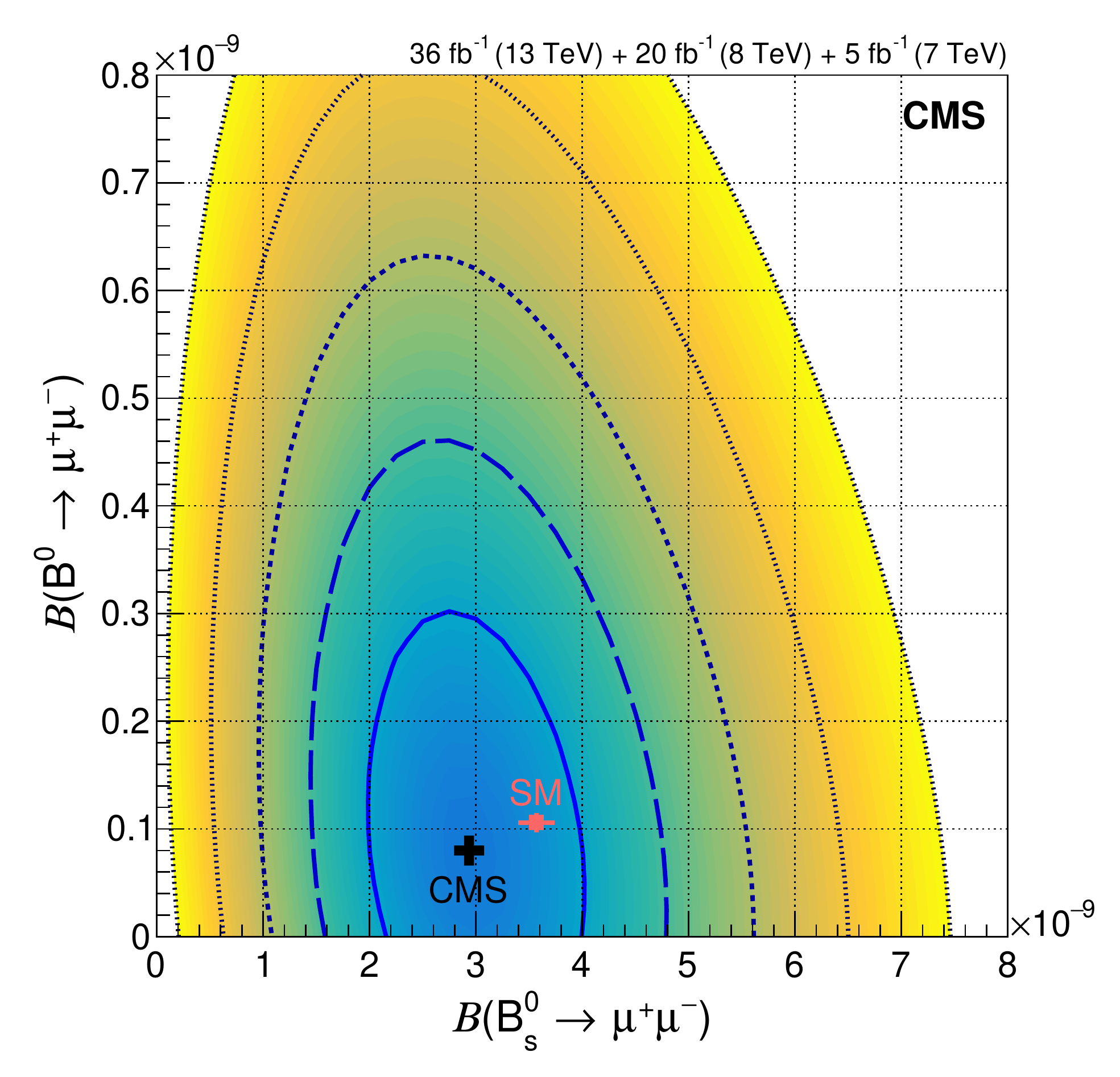}}
{\includegraphics[width=0.492\textwidth]{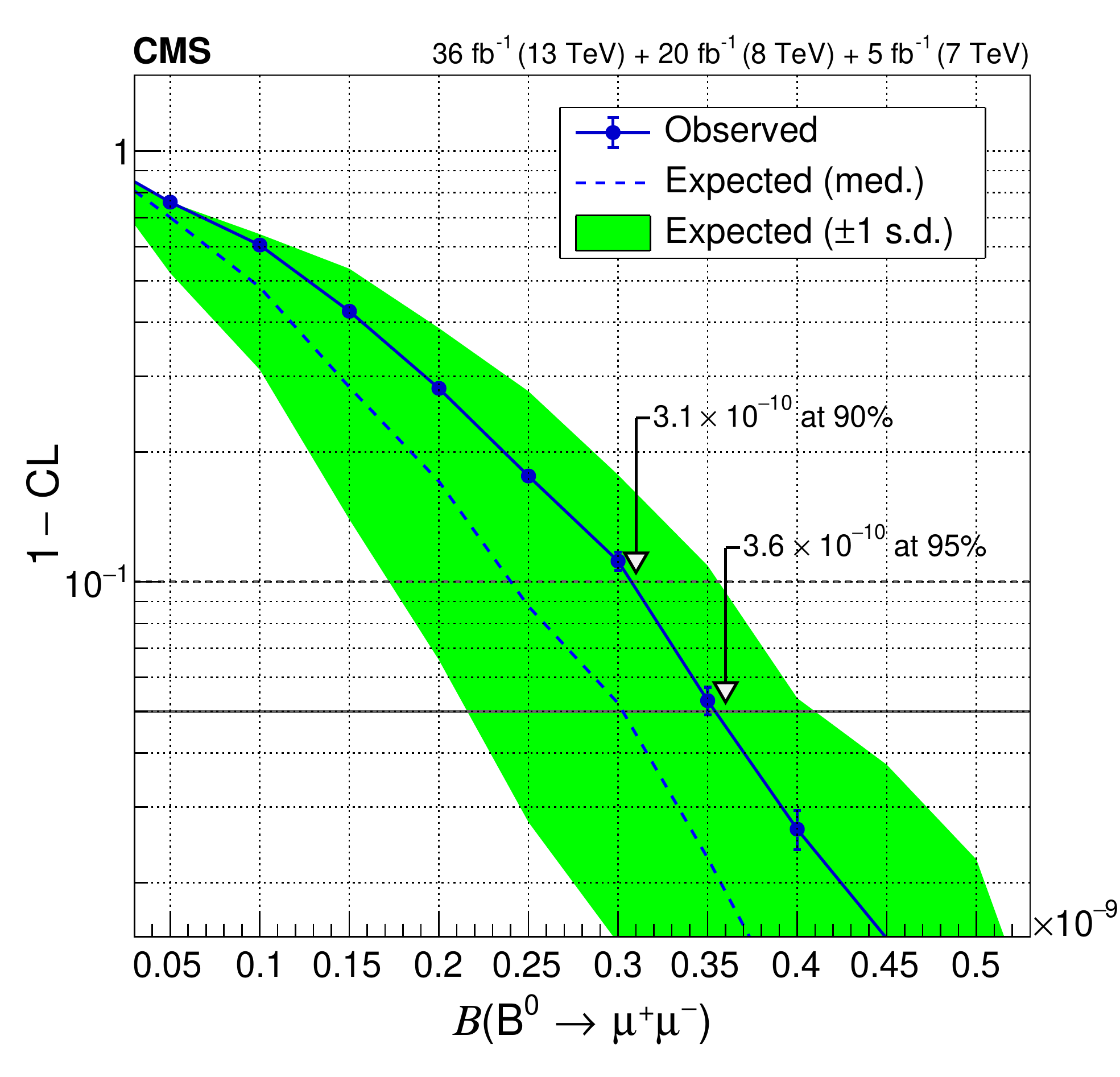}}
  \caption{(Left) Likelihood contours for the fit to the branching
    fractions $\cbf(\bsmm)$ and $\cbf(\bdmm)$, together with the
    best-fit value (cross) and the SM expectation (solid square). The contours correspond to regions with
    1--5 standard deviation coverage.  (Right) The quantity $1-\CL $ as a
    function of the assumed \bdmm\ branching fraction. The
    dashed curve shows the median expected value for the
    background-only hypothesis, while the solid line is
    the observed value. The shaded region indicates the $\pm1$ standard deviation
    uncertainty band. }
  \label{f:results}
\end{figure}

The likelihood contours of the fit are shown in Fig.~\ref{f:results}
(left), together with the SM expectation. The correlation coefficient
between the two branching fractions is $-0.181$.

Since no significant signal is observed for \bdmm, one-sided
upper limits are determined using the standard
\CLs\ rule~\cite{Read2002, Junk1999}, with the LHC-type profiled
likelihood as the test statistic. The result is $\cbf(\bdmm) <
\ulaBFBdmm\ (\ulbBFBdmm)$ at \ulacl\ (\ulbcl)\% confidence level
(\CL). The corresponding expected upper limit, assuming no signal, is
$\cbf(\bdmm) < \ulaExpBdmm\ (\ulbExpBdmm)$. In Fig.~\ref{f:results}
(right), the observed and expected confidence levels ($1-\CL$) are
shown versus the assumed \bdmm\ branching fraction.  In
  interpreting Fig.~\ref{f:results} (right), it is important to remember that
  the background-only hypothesis does not include the \bdmm\ signal
  expected in the SM.

All observed results are consistent within the uncertainties with the
SM expectations.
  Restricting the present analysis to the Run~1 data results in an
  observed branching fraction of $\cbf(\bsmm) = \resObsBFBsmmRunA$ with
  an observed (expected) significance of
  $\sigObsBFBsmmRunA$ ($\sigExpBFBsmmRunA$) standard deviations.
This is consistent with the results of Ref.~\cite{CMS:2014xfa}
when that analysis is restricted to the CMS data set.
The upper
limit on $\cbf(\bdmm)$ is substantially improved compared to the limit
of our previous study~\cite{Chatrchyan:2013bka}, even when using Run~1
data alone, because of more stringent muon identification criteria and
the introduction of the UML fit.

\begin{table}[!tb]
  \centering \topcaption{Summary of systematic uncertainty sources
    described in the text.  The uncertainties quoted for the branching
    fraction $\cbf(\bsmm)$ are relative uncertainties, while the
    uncertainties for the effective lifetime $\tmm$ are absolute and
    are given for both the 2D UML and \textit{sPlot} analysis methods. The
    relative uncertainties in the upper limit on $\cbf(\bdmm)$
    differ for the background yields and parametrization, but have negligible impact
    on that result. The bottom rows provide the total systematic
    uncertainty and the total uncertainty in the branching fraction
    and the effective lifetime measurements. Contributions that are
    included in other items are indicated by (*). Non-applicable sources
    are marked with dashes.}
  \label{t:sys}
  \begin{tabular}{llll}
    \hline
    Source\StrutTiny                  &$\cbf(\bsmm)$ [\%]  &\multicolumn{2}{c}{$\tmm$ [ps]} \\
    &        &2D UML  &\textit{sPlot}  \\
    \hline
    Kaon tracking                     &2.3--4  &\NA  &\NA      \\
    Normalization yield               &4       &\NA  &\NA      \\
    Background yields                 &1       &0.03 &(*)  \\
    Production process                &3       &\NA  &\NA      \\
    Muon identification               &3       &\NA  &\NA      \\
    Trigger                           &3       &\NA  &\NA      \\
    Efficiency (data/MC simulation)   &5--10   &\NA  &(*)    \\
    Efficiency (functional form)      &\NA     &0.01 &0.04 \\
    Efficiency lifetime dependence    &1--3    &(*)  &(*)  \\
    Running period dependence         &5--6    &0.07 &0.07 \\
    BDT discriminator threshold       &\NA     &0.02 &0.02 \\
    Silicon tracker alignment         &\NA     &0.02 &\NA    \\
    Finite size of MC sample          &\NA     &0.03 &\NA    \\
    Background parametrization/Fit bias  &2.3     &\NA  &0.09    \\
    \calc\ correction                  &\NA     &0.01 &0.01 \\[\cmsTabSkip]
    Absolute total systematic uncertainty\StrutTiny  &$0.3\times10^{-9}$ &0.09   &0.12 \\[\cmsTabSkip]
    Absolute total uncertainty\StrutTiny             &$0.7\times10^{-9}$ &$^{+0.61}_{-0.44}$  &$^{+0.52}_{-0.33}$ \\
   \hline
  \end{tabular}
\end{table}

\begin{table}[!tb]
  \centering
  \topcaption{Summary of the fitted yields for \bsmm, \bdmm, the
    combinatorial background for $5.2 < \mll < 5.45\GeV$, and the
    \bupsikp\ normalization, the average \pt\ of the \bsmm\ signal,
    and the ratio of efficiencies between the normalization and the
    signal for all 14 categories of the 3D UML branching fraction
    fit. The high and low ranges of the analysis BDT discriminator
    distribution are defined in Table~\ref{t:bfbdt}. The size of
    the peaking background is 5--10\% of the
    \bdmm\ signal. The average \pt\ is calculated from the MC
    simulation and has negligible uncertainties. The uncertainties shown
    include the statistical and systematic components. It should be
    noted that the \bsmm\ and \bdmm\ yields and their uncertainties  are determined from the branching
    fraction fit and  also include the normalization uncertainties.  }
  \label{t:bfyields}
  \begin{tabular}{lllclcl}
    \hline
    Category\StrutTiny           &$N(\Bs)$  &$N(\Bz)$  &$N_{\text{comb}}$  &$N_{\text{obs}}^{\Bp}/100$  &$\langle\pt(\Bs)\rangle
    [\GeV]$ &$\varepsilon_{\text{tot}}/\varepsilon_{\text{tot}}^{\Bp}$\\
    \hline
    \StrutTiny 2011/central    &\vuse{cat_2011s01_0_0_N_Bs} &\vuse{cat_2011s01_0_0_N_Bd}
    &\vuse{cat_2011s01_0_0_N_comb} &\vuse{cat_2011s01_0_0_N_Bu} &\vuse{cat_2011s01_0_0_ptBs}
    &\vuse{cat_2011s01_0_0_Eff_Bs_Bu}  \\
    2011/forward    &\vuse{cat_2011s01_1_0_N_Bs} &\vuse{cat_2011s01_1_0_N_Bd}
    &\vuse{cat_2011s01_1_0_N_comb} &\vuse{cat_2011s01_1_0_N_Bu} &\vuse{cat_2011s01_1_0_ptBs}
    &\vuse{cat_2011s01_1_0_Eff_Bs_Bu}  \\[\cmsTabSkip]
    \StrutTiny 2012/central/low    &\vuse{cat_2012s01_0_0_N_Bs} &\vuse{cat_2012s01_0_0_N_Bd}
    &\vuse{cat_2012s01_0_0_N_comb} &\vuse{cat_2012s01_0_0_N_Bu} &\vuse{cat_2012s01_0_0_ptBs}
    &\vuse{cat_2012s01_0_0_Eff_Bs_Bu}  \\
    2012/central/high   &\vuse{cat_2012s01_0_1_N_Bs} &\vuse{cat_2012s01_0_1_N_Bd}
    &\vuse{cat_2012s01_0_1_N_comb} &\vuse{cat_2012s01_0_1_N_Bu} &\vuse{cat_2012s01_0_1_ptBs}
    &\vuse{cat_2012s01_0_1_Eff_Bs_Bu}  \\
    2012/forward/low    &\vuse{cat_2012s01_1_0_N_Bs} &\vuse{cat_2012s01_1_0_N_Bd}
    &\vuse{cat_2012s01_1_0_N_comb} &\vuse{cat_2012s01_1_0_N_Bu} &\vuse{cat_2012s01_1_0_ptBs}
    &\vuse{cat_2012s01_1_0_Eff_Bs_Bu}  \\
    2012/forward/high   &\vuse{cat_2012s01_1_1_N_Bs} &\vuse{cat_2012s01_1_1_N_Bd}
    &\vuse{cat_2012s01_1_1_N_comb} &\vuse{cat_2012s01_1_1_N_Bu} &\vuse{cat_2012s01_1_1_ptBs}
    &\vuse{cat_2012s01_1_1_Eff_Bs_Bu}  \\[\cmsTabSkip]
    \StrutTiny 2016A/central/low    &\vuse{cat_2016BFs01_0_0_N_Bs} &\vuse{cat_2016BFs01_0_0_N_Bd}
    &\vuse{cat_2016BFs01_0_0_N_comb} &\vuse{cat_2016BFs01_0_0_N_Bu} &\vuse{cat_2016BFs01_0_0_ptBs}
    &\vuse{cat_2016BFs01_0_0_Eff_Bs_Bu}  \\
    2016A/central/high   &\vuse{cat_2016BFs01_0_1_N_Bs} &\vuse{cat_2016BFs01_0_1_N_Bd}
    &\vuse{cat_2016BFs01_0_1_N_comb} &\vuse{cat_2016BFs01_0_1_N_Bu} &\vuse{cat_2016BFs01_0_1_ptBs}
    &\vuse{cat_2016BFs01_0_1_Eff_Bs_Bu}  \\
    2016A/forward/low    &\vuse{cat_2016BFs01_1_0_N_Bs} &\vuse{cat_2016BFs01_1_0_N_Bd}
    &\vuse{cat_2016BFs01_1_0_N_comb} &\vuse{cat_2016BFs01_1_0_N_Bu} &\vuse{cat_2016BFs01_1_0_ptBs}
    &\vuse{cat_2016BFs01_1_0_Eff_Bs_Bu}  \\
    2016A/forward/high   &\vuse{cat_2016BFs01_1_1_N_Bs} &\vuse{cat_2016BFs01_1_1_N_Bd}
    &\vuse{cat_2016BFs01_1_1_N_comb} &\vuse{cat_2016BFs01_1_1_N_Bu} &\vuse{cat_2016BFs01_1_1_ptBs}
    &\vuse{cat_2016BFs01_1_1_Eff_Bs_Bu}  \\[\cmsTabSkip]
    \StrutTiny 2016B/central/low    &\vuse{cat_2016GHs01_0_0_N_Bs} &\vuse{cat_2016GHs01_0_0_N_Bd}
    &\vuse{cat_2016GHs01_0_0_N_comb} &\vuse{cat_2016GHs01_0_0_N_Bu} &\vuse{cat_2016GHs01_0_0_ptBs}
    &\vuse{cat_2016GHs01_0_0_Eff_Bs_Bu}  \\
    2016B/central/high   &\vuse{cat_2016GHs01_0_1_N_Bs} &\vuse{cat_2016GHs01_0_1_N_Bd}
    &\vuse{cat_2016GHs01_0_1_N_comb} &\vuse{cat_2016GHs01_0_1_N_Bu} &\vuse{cat_2016GHs01_0_1_ptBs}
    &\vuse{cat_2016GHs01_0_1_Eff_Bs_Bu}  \\
    2016B/forward/low    &\vuse{cat_2016GHs01_1_0_N_Bs} &\vuse{cat_2016GHs01_1_0_N_Bd}
    &\vuse{cat_2016GHs01_1_0_N_comb} &\vuse{cat_2016GHs01_1_0_N_Bu} &\vuse{cat_2016GHs01_1_0_ptBs}
    &\vuse{cat_2016GHs01_1_0_Eff_Bs_Bu}  \\
    2016B/forward/high   &\vuse{cat_2016GHs01_1_1_N_Bs} &\vuse{cat_2016GHs01_1_1_N_Bd}
    &\vuse{cat_2016GHs01_1_1_N_comb} &\vuse{cat_2016GHs01_1_1_N_Bu} &\vuse{cat_2016GHs01_1_1_ptBs}
    &\vuse{cat_2016GHs01_1_1_Eff_Bs_Bu}  \\[\cmsTabSkip]
    \hline
  \end{tabular}
\end{table}

\section{Effective lifetime measurement}
\label{s:lifetime}
Two independent fitting methods were developed to determine the
\bsmm\ effective lifetime, in order to allow for extensive validation,
cross-checks, and systematic studies. The primary method consists of
an extended 2D UML fit to the dimuon invariant mass and proper decay
time~\cite{Chatrchyan:2013sxa, Sirunyan:2017nbv}.  The second method employs a 1D approach
in which the background is subtracted using
the \textit{sPlot}~\cite{Pivk:2004ty} technique, with a function then
fitted to the background-subtracted distribution using a binned ML.
The 2D UML approach was chosen as the primary method, prior to
analyzing the events in the signal region, because it exhibited better median expected performance.

A slightly modified analysis BDT setup is used for both
approaches. For simplicity, a single
BDT discriminator requirement, as indicated in Table~\ref{t:tau1bdt},
is used for each channel. As before, these
requirements are determined by optimizing the expected performance.

\begin{table}[!tb]
  \centering
  \topcaption{Analysis BDT discriminator minimum requirements per
    channel and running period for the 1D and 2D effective lifetime fits.}
  \label{t:tau1bdt}
  \begin{tabular}{cccccccc}
    \hline
    \multicolumn{2}{c}{2011}&\multicolumn{2}{c}{2012}&\multicolumn{2}{c}{2016A}&\multicolumn{2}{c}{2016B}\\
    Central&Forward&Central&Forward&Central&Forward&Central&Forward\\
    \hline
    0.22 &0.19
    &0.32 &0.32
    &0.22 &0.30
    &0.22 &0.29
    \\
    \hline
  \end{tabular}
\end{table}

The fits are performed over the decay time range $1 < t < 11\ps$.
For very short times, the reconstruction efficiency is small because
of the flight-length significance and isolation requirements, while
for long times, the efficiency is reduced in Run 2 because of the HLT requirements.
The results of both approaches are limited by the small
number of signal events. Systematic uncertainties have only a small
impact on the total uncertainty.

\subsection{Two-dimensional unbinned maximum likelihood fit}
\label{s:tau2}
The extended 2D UML fit uses the proper decay time resolution, \st, as a
conditional parameter.
The unnormalized PDF $\mathcal{P}$ has the following expression:
\begin{linenomath}
  \begin{equation}
    \begin{split}
      \mathcal{P}(\mll, t; \st) = N_{\text{sig}} P_{\text{sig}}(\mll) T_{\text{sig}}(t; \st) \varepsilon_{\text{sig}}(t) + N_{\text{comb}} P_{\text{comb}}(\mll) T_{\text{comb}}(t; \st) \\
      + N_{\text{peak}} P_{\text{peak}}(\mll) T_{\text{peak}}(t; \st)\varepsilon_{\text{peak}}(t) + N_{\text{semi}} P_{\text{semi}}(\mll) T_{\text{semi}}(t; \st)\varepsilon_{\text{semi}}(t),
    \end{split}
  \end{equation}
\end{linenomath}
where $N_{\text{sig}}$, $N_{\text{comb}}$, $N_{\text{peak}}$, and
$N_{\text{semi}}$ are the \bsmm\ signal, combinatorial, peaking (both
\bdmm\ and \bhh), and semileptonic (\bhm\ combined with \bhmm)
background yield Poisson terms, respectively. The invariant mass and
decay time PDFs for the signal and background components are described
by $P(\mll)$ and $T(t; \st)$, respectively. The efficiencies
$\varepsilon_{\text{sig}}$, $\varepsilon_{\text{peak}}$,
and $\varepsilon_{\text{semi}}$ for the signal and background components
as a function of the proper decay time are determined from simulated
event samples.

For the PDFs describing the dimuon invariant mass distribution
($P_{\text{sig}}, P_{\text{comb}}, P_{\text{peak}}, P_{\text{semi}}$),
the signal shape is parametrized by a Crystal Ball function, while the
combinatorial, peaking, and semileptonic background contributions are
parametrized by a nonnegative Bernstein polynomial of the first
degree, the sum of a Crystal Ball function and a Gaussian function with common
mean, and a Gaussian function, respectively. For the PDFs describing the proper
decay time distributions ($T_{\text{sig}}, T_{\text{comb}},
T_{\text{peak}}, T_{\text{semi}}$), the signal and background shapes
are parametrized by individual exponential functions, which are
convolved with a Gaussian function to incorporate the effect of the detector
resolution, and with the $T_{\text{sig}}$ exponential function depending on the effective lifetime.
The signal, peaking, and semileptonic background proper
decay time PDFs are corrected with their respective efficiency factors
($\varepsilon_{\text{sig}}, \varepsilon_{\text{peak}},
\varepsilon_{\text{semi}}$).
Such a correction is not applied to the combinatorial background since the
decay time PDF $T_{\text{comb}}$ is modeled from data directly.
The modeling of the efficiency function
has a systematic uncertainty of about $0.01\ps$, determined by
variation of the parametrization.

The signal yield, the effective lifetime \tmm, and all parameters
of the combinatorial background (except the combinatorial
background in the forward channel of 2011, which is held fixed
because there are no events in the sideband) are determined in the fit. All other
parameters are either constrained (background yields of the peaking
and semileptonic components inside log-normal constraints) or fixed to
the MC simulation values (all other parameters).  The UML fit is
performed simultaneously in the eight independent channels defined in
Table~\ref{t:tau1bdt}.

\begin{figure}[!tb]
  \centering
{\includegraphics[width=0.492\textwidth]{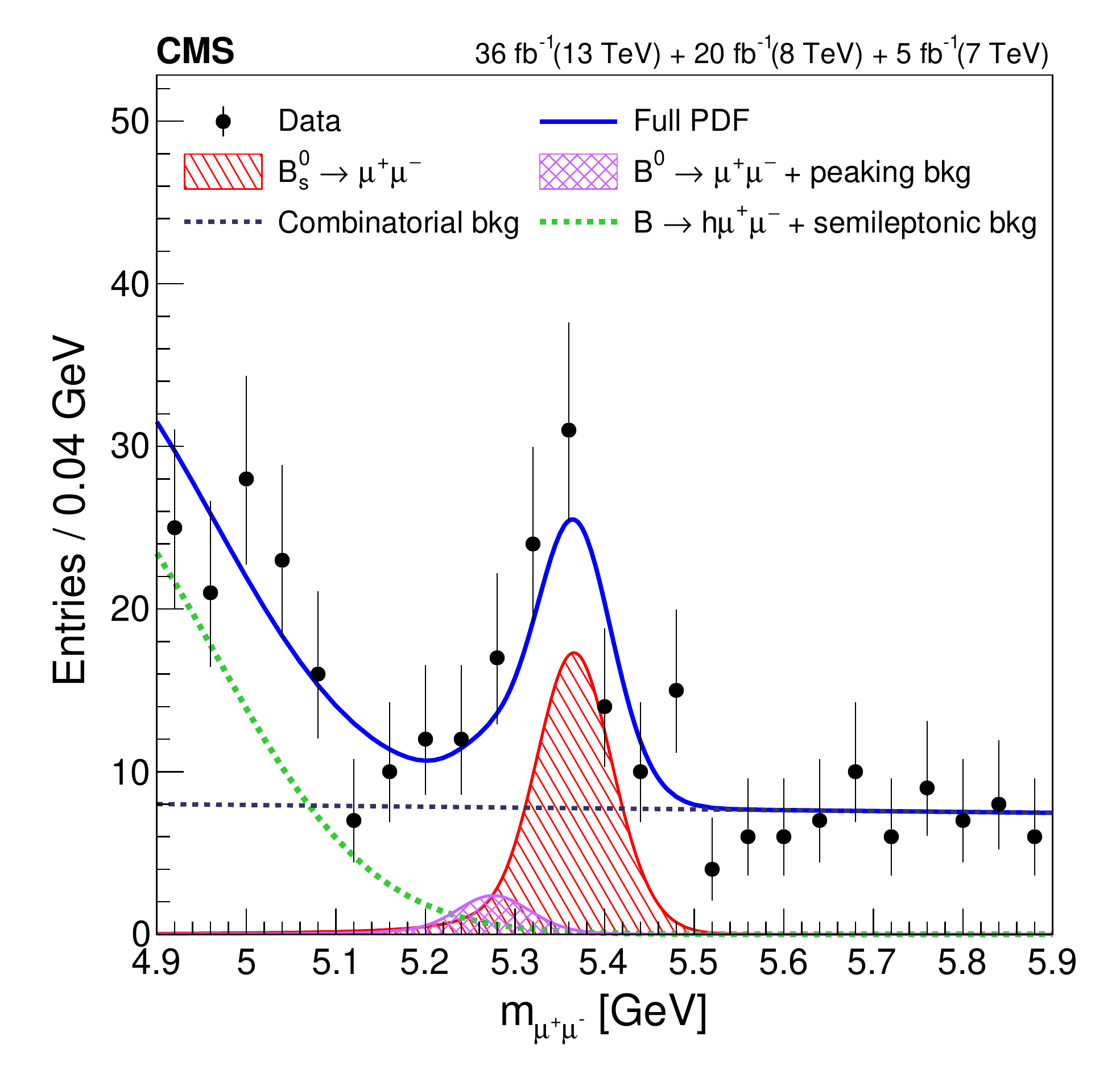}}
{\includegraphics[width=0.492\textwidth]{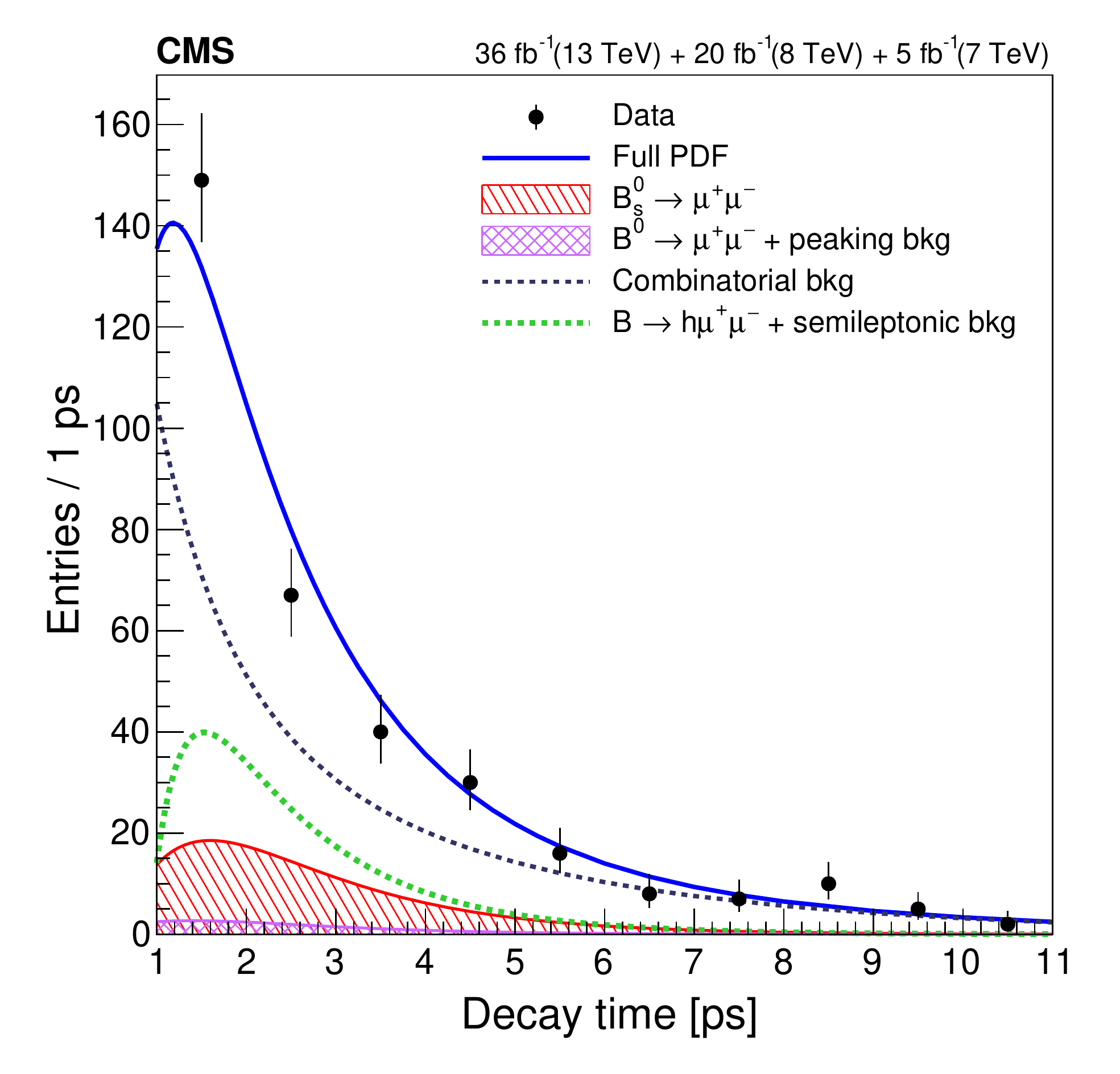}}
  \caption{Invariant mass (left) and proper decay time (right)
    distributions, with the 2D UML fit projections overlaid. The data
    combine all channels passing the analysis BDT discriminator requirements as
    given in Table~\ref{t:tau1bdt}. The total fit is shown by the
    solid line and the different background components by the broken lines
    and cross-hatched distributions. The signal component is shown
    by the single-hatched distribution.  }
  \label{f:teff2d}
\end{figure}

The combined  mass and proper decay time distributions from all
channels, with the fit results overlaid, are shown in Fig.~\ref{f:teff2d}.
The effective lifetime obtained from the fit is
\begin{linenomath}
  \begin{equation}
    \tmm = \resObsTauBsmm\ps,
  \end{equation}
\end{linenomath}
where the uncertainty represents the combined statistical and
systematic terms. Systematic uncertainties, beyond those already
discussed, include small contributions from the limited statistical
precision of the MC simulation ($0.03\ps$) and the \calc\ correction ($0.01\ps$).
All systematic uncertainties are summarized in Table~\ref{t:sys}.

\subsection{One-dimensional binned maximum likelihood fit}
\label{s:tau1}

In the second method, the complete PDF described in Section~\ref{s:bf}
is used to determine \textit{sPlot} weights.  All events passing the
analysis BDT requirements in Table~\ref{t:tau1bdt}, but without any
proper decay time selection, are used for this step. Because of the
small number of events in individual channels, an integration is
performed over the central and forward channels for both the Run~1 and
Run~2 data. The effective lifetime is determined with an exponential
function modified to include the channel-dependent resolution and
efficiency effects. To properly determine the uncertainty in the
effective lifetime from the weighted fit, a custom
algorithm~\cite{James:1019859} is implemented. This algorithm has
several features. First, it performs a weighted binned ML fit to the
\textit{sPlot} distribution to provide the correct central value and
covariance matrix. Second, to reduce biases associated with large
histogram bin widths, it calculates the integral of the PDF,
integrating over bins. Third, it incorporates a resolution and
efficiency model into the effective decay time PDF. Finally, it
provides asymmetric uncertainties in the fit parameters.  The
determined effective lifetime and the associated variance are
consistent with the expectations from pseudo-experiments and the
statistical uncertainties are in agreement with the confidence
intervals reported by the Neyman construction~\cite{Neyman:1937uhy}.

\begin{figure}[!tbp]
  \centering
{\includegraphics[width=0.492\textwidth]{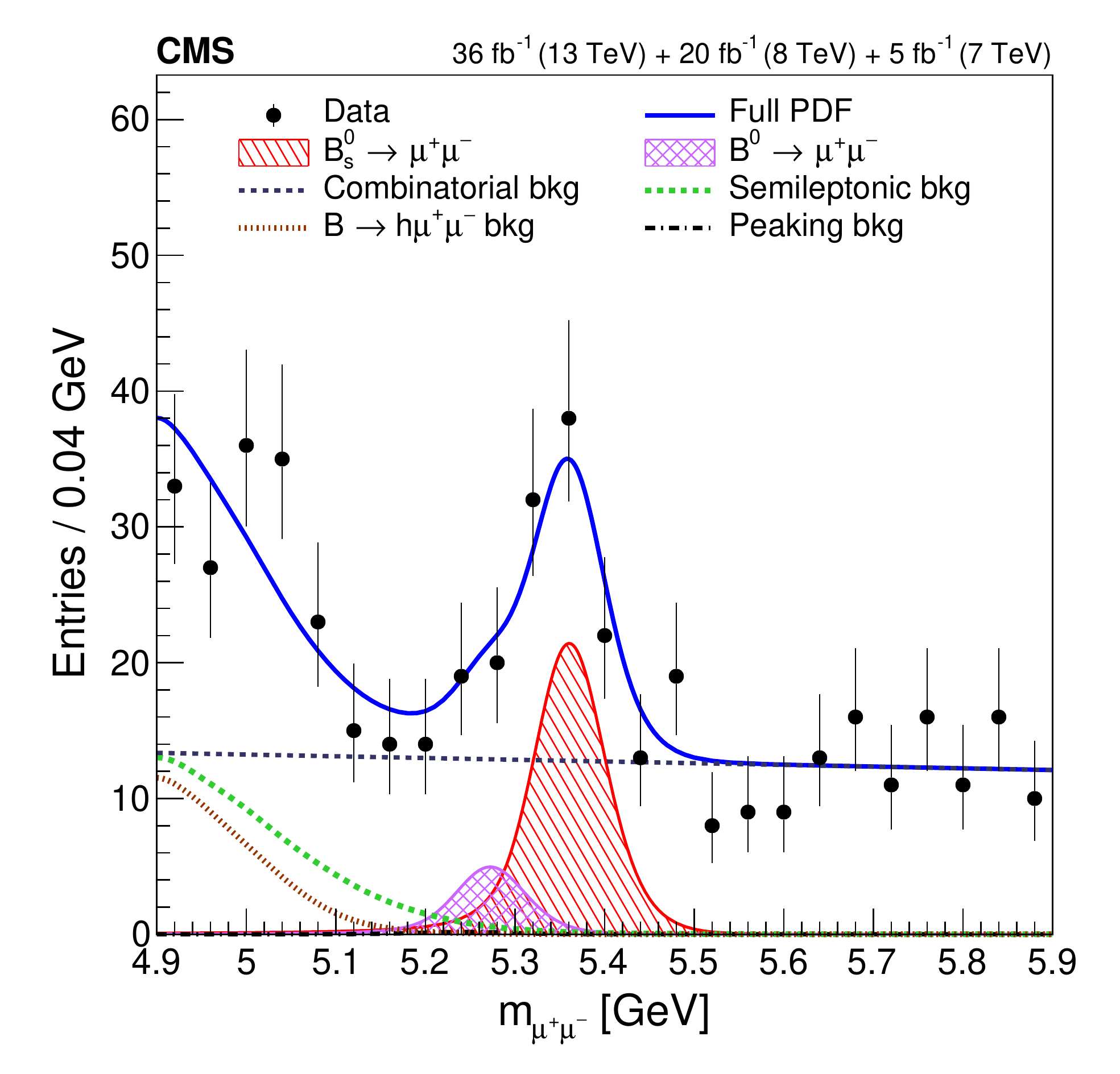}}
{\includegraphics[width=0.492\textwidth]{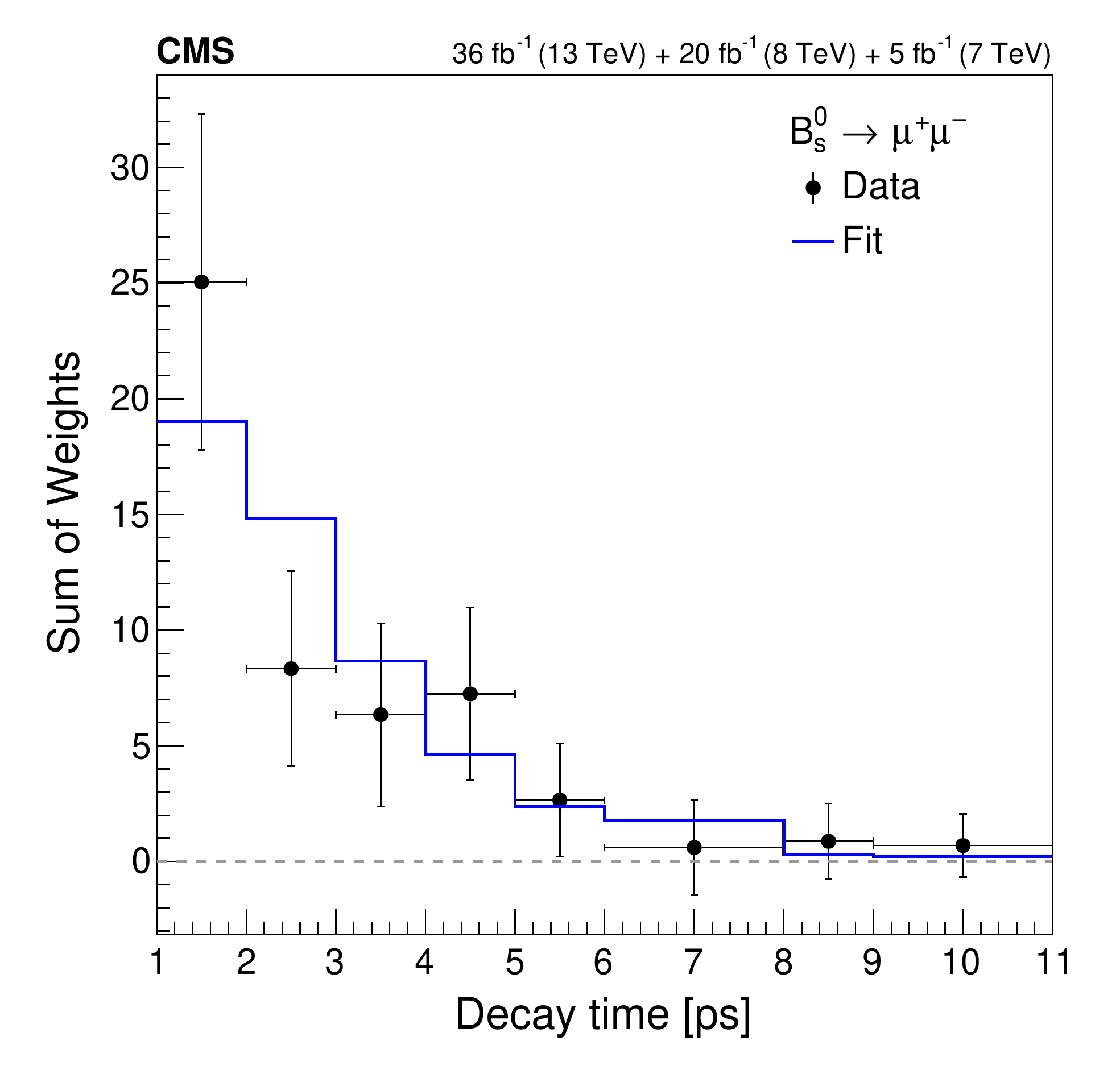}}
  \caption{Invariant mass (left) and proper decay time (right)
    distributions, with the \textit{sPlot} fit projections overlaid. The data
    combine all channels passing the analysis BDT discriminator requirements as
    given in Table~\ref{t:tau1bdt}. For the mass distribution, no
    requirement on the decay time is applied. The total fit is shown by the
    solid line, the different background components by the broken lines
    and cross-hatched distribution. The signal component is shown
    by the single-hatched distribution. }
  \label{f:splotlifetime}
\end{figure}

Figure~\ref{f:splotlifetime} shows the mass distribution of all
contributing data, without requiring $t > 1\ps$, and the weighted
signal proper decay time distribution, together with the result of the
binned ML fit. The fit yields $\tmm = \resObsTauSplot\ps$,
where the uncertainty is the combination of the statistical and systematic contributions.
Using pseudo-experiments with post-fit nuisance
  parameters, a fit bias of $+0.09\ps$ is observed and corrected for
  in the result above. It is included as a systematic uncertainty. The
  reasons for this bias are, first, negative yields are not allowed in
  the weighted ML fit and, second, the sample size at large decay times is very small.  The
  decay time dependence of the selection efficiency leads to a
  systematic uncertainty of $0.04\ps$. All systematic uncertainties
are summarized in Table~\ref{t:sys}.

The two fitting methods, the 2D UML fit and the 1D \textit{sPlot}
approach, yield consistent results. The observed total uncertainties
in the primary fitting method are about one root-mean-square deviation
larger than the expected median uncertainties ($^{+0.39}_{-0.30}\ps$).
The expected median uncertainty for the 1D \textit{sPlot} approach are
$^{+0.49}_{-0.31}\ps$. While the uncertainties are sizable, the
results are consistent with the SM expectation that only the heavy
$\B_{s\text{H}}$ state contributes to the \bsmm\ decay.

\section{Summary}
\label{s:summary}
Measurements of the rare leptonic $\B$ meson decays \bsmm\ and
\bdmm\ have been performed in $\p\p$ collision data collected by the CMS
experiment at the LHC, corresponding to integrated luminosities of
5\fbinv at center-of-mass energy 7\TeV, 20\fbinv at 8\TeV, and
36\fbinv at 13\TeV.
The \bsmm\ decay is observed with a significance of $\sigObsBFBsmm$
standard deviations and the time-integrated branching fraction is
measured to be $\cbf(\bsmm) = \resObsBFBsmmLong$, where the last
uncertainty refers to the uncertainty in the ratio
of the \PBzs\ and the \PBp\ fragmentation
functions. No significant
\bdmm\ signal is observed and an upper limit $\cbf(\bdmm) <
\ulaBFBdmm$ is determined at \ulacl\% confidence level. The
\bsmm\ effective lifetime is found to be $\tmm = \resObsTauBsmmLong\ps$.  The
results for the branching fractions supersede the previous results from CMS~\cite{Chatrchyan:2013bka},
which were based on the 7 and 8\TeV\ data only. All of the results are in agreement with
the standard model predictions.

\begin{acknowledgments}
We congratulate our colleagues in the CERN accelerator departments for the excellent performance of the LHC and thank the technical and administrative staffs at CERN and at other CMS institutes for their contributions to the success of the CMS effort. In addition, we gratefully acknowledge the computing centers and personnel of the Worldwide LHC Computing Grid for delivering so effectively the computing infrastructure essential to our analyses. Finally, we acknowledge the enduring support for the construction and operation of the LHC and the CMS detector provided by the following funding agencies: BMBWF and FWF (Austria); FNRS and FWO (Belgium); CNPq, CAPES, FAPERJ, FAPERGS, and FAPESP (Brazil); MES (Bulgaria); CERN; CAS, MoST, and NSFC (China); COLCIENCIAS (Colombia); MSES and CSF (Croatia); RPF (Cyprus); SENESCYT (Ecuador); MoER, ERC IUT, PUT and ERDF (Estonia); Academy of Finland, MEC, and HIP (Finland); CEA and CNRS/IN2P3 (France); BMBF, DFG, and HGF (Germany); GSRT (Greece); NKFIA (Hungary); DAE and DST (India); IPM (Iran); SFI (Ireland); INFN (Italy); MSIP and NRF (Republic of Korea); MES (Latvia); LAS (Lithuania); MOE and UM (Malaysia); BUAP, CINVESTAV, CONACYT, LNS, SEP, and UASLP-FAI (Mexico); MOS (Montenegro); MBIE (New Zealand); PAEC (Pakistan); MSHE and NSC (Poland); FCT (Portugal); JINR (Dubna); MON, RosAtom, RAS, RFBR, and NRC KI (Russia); MESTD (Serbia); SEIDI, CPAN, PCTI, and FEDER (Spain); MOSTR (Sri Lanka); Swiss Funding Agencies (Switzerland); MST (Taipei); ThEPCenter, IPST, STAR, and NSTDA (Thailand); TUBITAK and TAEK (Turkey); NASU (Ukraine); STFC (United Kingdom); DOE and NSF (USA).

\hyphenation{Rachada-pisek} Individuals have received support from the Marie-Curie program and the European Research Council and Horizon 2020 Grant, contract Nos.\ 675440, 752730, and 765710 (European Union); the Leventis Foundation; the A.P.\ Sloan Foundation; the Alexander von Humboldt Foundation; the Belgian Federal Science Policy Office; the Fonds pour la Formation \`a la Recherche dans l'Industrie et dans l'Agriculture (FRIA-Belgium); the Agentschap voor Innovatie door Wetenschap en Technologie (IWT-Belgium); the F.R.S.-FNRS and FWO (Belgium) under the ``Excellence of Science -- EOS" -- be.h project n.\ 30820817; the Beijing Municipal Science \& Technology Commission, No. Z181100004218003; the Ministry of Education, Youth and Sports (MEYS) of the Czech Republic; the Lend\"ulet (``Momentum") Program and the J\'anos Bolyai Research Scholarship of the Hungarian Academy of Sciences, the New National Excellence Program \'UNKP, the NKFIA research grants 123842, 123959, 124845, 124850, 125105, 128713, 128786, and 129058 (Hungary); the Council of Science and Industrial Research, India; the HOMING PLUS program of the Foundation for Polish Science, cofinanced from European Union, Regional Development Fund, the Mobility Plus program of the Ministry of Science and Higher Education, the National Science Center (Poland), contracts Harmonia 2014/14/M/ST2/00428, Opus 2014/13/B/ST2/02543, 2014/15/B/ST2/03998, and 2015/19/B/ST2/02861, Sonata-bis 2012/07/E/ST2/01406; the National Priorities Research Program by Qatar National Research Fund; the Ministry of Science and Education, grant no. 3.2989.2017 (Russia); the Programa Estatal de Fomento de la Investigaci{\'o}n Cient{\'i}fica y T{\'e}cnica de Excelencia Mar\'{\i}a de Maeztu, grant MDM-2015-0509 and the Programa Severo Ochoa del Principado de Asturias; the Thalis and Aristeia programs cofinanced by EU-ESF and the Greek NSRF; the Rachadapisek Sompot Fund for Postdoctoral Fellowship, Chulalongkorn University and the Chulalongkorn Academic into Its 2nd Century Project Advancement Project (Thailand); the Nvidia Corporation; the Welch Foundation, contract C-1845; and the Weston Havens Foundation (USA).
\end{acknowledgments}

\bibliography{auto_generated}
\cleardoublepage \appendix\section{The CMS Collaboration \label{app:collab}}\begin{sloppypar}\hyphenpenalty=5000\widowpenalty=500\clubpenalty=5000\vskip\cmsinstskip
\textbf{Yerevan Physics Institute, Yerevan, Armenia}\\*[0pt]
A.M.~Sirunyan$^{\textrm{\dag}}$, A.~Tumasyan
\vskip\cmsinstskip
\textbf{Institut f\"{u}r Hochenergiephysik, Wien, Austria}\\*[0pt]
W.~Adam, F.~Ambrogi, T.~Bergauer, J.~Brandstetter, M.~Dragicevic, J.~Er\"{o}, A.~Escalante~Del~Valle, M.~Flechl, R.~Fr\"{u}hwirth\cmsAuthorMark{1}, M.~Jeitler\cmsAuthorMark{1}, N.~Krammer, I.~Kr\"{a}tschmer, D.~Liko, T.~Madlener, I.~Mikulec, N.~Rad, J.~Schieck\cmsAuthorMark{1}, R.~Sch\"{o}fbeck, M.~Spanring, D.~Spitzbart, W.~Waltenberger, C.-E.~Wulz\cmsAuthorMark{1}, M.~Zarucki
% ...
\vskip\cmsinstskip
\dag: Deceased\\
1:  Also at Vienna University of Technology, Vienna, Austria\\
% ...
\end{sloppypar}
\end{document}